\documentclass[%
 reprint,
 superscriptaddress,
 nofootinbib,
 amsmath,amssymb,
 aps,
 prd,
]{revtex4-2}

\usepackage{xcolor}
\usepackage{graphicx}
\usepackage{dcolumn}
\usepackage{bm}
\usepackage{hyperref}
\usepackage{fontawesome5}
\usepackage{caption}
\usepackage{subcaption}
\usepackage{orcidlink}
\usepackage{appendix}

\usepackage{soul}

\DeclareMathOperator{\diag}{diag}

\definecolor{mpl_blue}{HTML}{1F77B4}
\definecolor{mpl_orange}{HTML}{FF7F0E}
\definecolor{mpl_green}{HTML}{2CA02C}
\definecolor{mpl_red}{HTML}{D62728}

\newcommand{\github}[1]{%
   \href{#1}{\faGithubSquare}%
}
\newcommand{\vect}[1]{\boldsymbol{#1}}

\begin{document}

\preprint{APS/123-QED}

\title{The NANOGrav 15-year Gravitational-Wave Background Methods}

\author{Aaron D. Johnson\,\orcidlink{0000-0002-7445-8423}}
\affiliation{Center for Gravitation, Cosmology and Astrophysics, Department of Physics, University of Wisconsin-Milwaukee,\\ P.O. Box 413, Milwaukee, WI 53201, USA}
\affiliation{Division of Physics, Mathematics, and Astronomy, California Institute of Technology, Pasadena, CA 91125, USA}
\author{Patrick M. Meyers\,\orcidlink{0000-0002-2689-0190}}
\affiliation{Division of Physics, Mathematics, and Astronomy, California Institute of Technology, Pasadena, CA 91125, USA}
\author{Paul T. Baker\,\orcidlink{0000-0003-2745-753X}}
\affiliation{Department of Physics and Astronomy, Widener University, One University Place, Chester, PA 19013, USA}
\author{Neil J. Cornish\,\orcidlink{0000-0002-7435-0869}}
\affiliation{Department of Physics, Montana State University, Bozeman, MT 59717, USA}
\author{Jeffrey S. Hazboun\,\orcidlink{0000-0003-2742-3321}}
\affiliation{Department of Physics, Oregon State University, Corvallis, OR 97331, USA}
\author{Tyson B. Littenberg\,\orcidlink{0000-0002-9574-578X}}
\affiliation{NASA Marshall Space Flight Center, Huntsville, AL 35812, USA}
\author{Joseph D. Romano\,\orcidlink{0000-0003-4915-3246}}
\affiliation{Department of Physics, Texas Tech University, Box 41051, Lubbock, TX 79409, USA}
\author{Stephen R. Taylor\,\orcidlink{0000-0003-0264-1453}}
\affiliation{Department of Physics and Astronomy, Vanderbilt University, 2301 Vanderbilt Place, Nashville, TN 37235, USA}
\author{Michele Vallisneri\,\orcidlink{0000-0002-4162-0033}}
\affiliation{Jet Propulsion Laboratory, California Institute of Technology, 4800 Oak Grove Drive, Pasadena, CA 91109, USA}
\affiliation{Division of Physics, Mathematics, and Astronomy, California Institute of Technology, Pasadena, CA 91125, USA}
\author{Sarah J. Vigeland\,\orcidlink{0000-0003-4700-9072}}
\affiliation{Center for Gravitation, Cosmology and Astrophysics, Department of Physics, University of Wisconsin-Milwaukee,\\ P.O. Box 413, Milwaukee, WI 53201, USA}
\author{Ken D. Olum\,\orcidlink{0000-0002-2027-3714}}
\affiliation{Institute of Cosmology, Department of Physics and Astronomy, Tufts University, Medford, MA 02155, USA}
\author{Xavier Siemens\,\orcidlink{0000-0002-7778-2990}}
\affiliation{Department of Physics, Oregon State University, Corvallis, OR 97331, USA}
\affiliation{Center for Gravitation, Cosmology and Astrophysics, Department of Physics, University of Wisconsin-Milwaukee,\\ P.O. Box 413, Milwaukee, WI 53201, USA}
\author{Justin A. Ellis}
\affiliation{Department of Astronomy, University of Maryland, College Park, MD 20742}
\affiliation{Center for Research and Exploration in Space Science and Technology, NASA/GSFC, Greenbelt, MD 20771}
\affiliation{NASA Goddard Space Flight Center, Greenbelt, MD 20771, USA}
\altaffiliation{Infinia ML, 202 Rigsbee Avenue, Durham NC, 27701}
\author{Rutger van Haasteren\,\orcidlink{0000-0002-6428-2620}}
\affiliation{Max-Planck-Institut f{\"u}r Gravitationsphysik (Albert-Einstein-Institut), Callinstrasse 38, D-30167, Hannover, Germany}
\author{Sophie Hourihane\,\orcidlink}
\affiliation{Division of Physics, Mathematics, and Astronomy, California Institute of Technology, Pasadena, CA 91125, USA}
\author{Gabriella Agazie\,\orcidlink{0000-0001-5134-3925}}
\affiliation{Center for Gravitation, Cosmology and Astrophysics, Department of Physics, University of Wisconsin-Milwaukee,\\ P.O. Box 413, Milwaukee, WI 53201, USA}
\author{Akash Anumarlapudi\,\orcidlink{0000-0002-8935-9882}}
\affiliation{Center for Gravitation, Cosmology and Astrophysics, Department of Physics, University of Wisconsin-Milwaukee,\\ P.O. Box 413, Milwaukee, WI 53201, USA}
\author{Anne M. Archibald\,\orcidlink{0000-0003-0638-3340}}
\affiliation{Newcastle University, NE1 7RU, UK}
\author{Zaven Arzoumanian\,\orcidlink}
\affiliation{X-Ray Astrophysics Laboratory, NASA Goddard Space Flight Center, Code 662, Greenbelt, MD 20771, USA}
\author{Laura Blecha\,\orcidlink{0000-0002-2183-1087}}
\affiliation{Physics Department, University of Florida, Gainesville, FL 32611, USA}
\author{Adam Brazier\,\orcidlink{0000-0001-6341-7178}}
\affiliation{Cornell Center for Astrophysics and Planetary Science and Department of Astronomy, Cornell University, Ithaca, NY 14853, USA}
\affiliation{Cornell Center for Advanced Computing, Cornell University, Ithaca, NY 14853, USA}
\author{Paul R. Brook\,\orcidlink{0000-0003-3053-6538}}
\affiliation{Institute for Gravitational Wave Astronomy and School of Physics and Astronomy, University of Birmingham, Edgbaston, Birmingham B15 2TT, UK}
\author{Sarah Burke-Spolaor\,\orcidlink{0000-0003-4052-7838}}
\affiliation{Department of Physics and Astronomy, West Virginia University, P.O. Box 6315, Morgantown, WV 26506, USA}
\affiliation{Center for Gravitational Waves and Cosmology, West Virginia University, Chestnut Ridge Research Building, Morgantown, WV 26505, USA}
\author{Bence B\'{e}csy\,\orcidlink{0000-0003-0909-5563}}
\affiliation{Department of Physics, Oregon State University, Corvallis, OR 97331, USA}
\author{J. Andrew Casey-Clyde\,\orcidlink{0000-0002-5557-4007}}
\affiliation{Department of Physics, University of Connecticut, 196 Auditorium Road, U-3046, Storrs, CT 06269-3046, USA}
\author{Maria Charisi\,\orcidlink{0000-0003-3579-2522}}
\affiliation{Department of Physics and Astronomy, Vanderbilt University, 2301 Vanderbilt Place, Nashville, TN 37235, USA}
\author{Shami Chatterjee\,\orcidlink{0000-0002-2878-1502}}
\affiliation{Cornell Center for Astrophysics and Planetary Science and Department of Astronomy, Cornell University, Ithaca, NY 14853, USA}
\author{Katerina Chatziioannou\,\orcidlink}
\affiliation{Division of Physics, Mathematics, and Astronomy, California Institute of Technology, Pasadena, CA 91125, USA}
\author{Tyler Cohen\,\orcidlink{0000-0001-7587-5483}}
\affiliation{Department of Physics, New Mexico Institute of Mining and Technology, 801 Leroy Place, Socorro, NM 87801, USA}
\author{James M. Cordes\,\orcidlink{0000-0002-4049-1882}}
\affiliation{Cornell Center for Astrophysics and Planetary Science and Department of Astronomy, Cornell University, Ithaca, NY 14853, USA}
\author{Fronefield Crawford\,\orcidlink{0000-0002-2578-0360}}
\affiliation{Department of Physics and Astronomy, Franklin \& Marshall College, P.O. Box 3003, Lancaster, PA 17604, USA}
\author{H. Thankful Cromartie\,\orcidlink{0000-0002-6039-692X}}
\altaffiliation{NASA Hubble Fellowship: Einstein Postdoctoral Fellow}
\affiliation{Cornell Center for Astrophysics and Planetary Science and Department of Astronomy, Cornell University, Ithaca, NY 14853, USA}
\author{Kathryn Crowter\,\orcidlink{0000-0002-1529-5169}}
\affiliation{Department of Physics and Astronomy, University of British Columbia, 6224 Agricultural Road, Vancouver, BC V6T 1Z1, Canada}
\author{Megan E. DeCesar\,\orcidlink{0000-0002-2185-1790}}
\affiliation{George Mason University, resident at the Naval Research Laboratory, Washington, DC 20375, USA}
\author{Paul B. Demorest\,\orcidlink{0000-0002-6664-965X}}
\affiliation{National Radio Astronomy Observatory, 1003 Lopezville Rd., Socorro, NM 87801, USA}
\author{Timothy Dolch\,\orcidlink{0000-0001-8885-6388}}
\affiliation{Department of Physics, Hillsdale College, 33 E. College Street, Hillsdale, MI 49242, USA}
\affiliation{Eureka Scientific, 2452 Delmer Street, Suite 100, Oakland, CA 94602-3017, USA}
\author{Brendan Drachler\,\orcidlink}
\affiliation{School of Physics and Astronomy, Rochester Institute of Technology, Rochester, NY 14623, USA}
\affiliation{Laboratory for Multiwavelength Astrophysics, Rochester Institute of Technology, Rochester, NY 14623, USA}
\author{Elizabeth C. Ferrara\,\orcidlink{0000-0001-7828-7708}}
\affiliation{Department of Astronomy, University of Maryland, College Park, MD 20742}
\affiliation{Center for Research and Exploration in Space Science and Technology, NASA/GSFC, Greenbelt, MD 20771}
\affiliation{NASA Goddard Space Flight Center, Greenbelt, MD 20771, USA}
\author{William Fiore\,\orcidlink{0000-0001-5645-5336}}
\affiliation{Department of Physics and Astronomy, West Virginia University, P.O. Box 6315, Morgantown, WV 26506, USA}
\affiliation{Center for Gravitational Waves and Cosmology, West Virginia University, Chestnut Ridge Research Building, Morgantown, WV 26505, USA}
\author{Emmanuel Fonseca\,\orcidlink{0000-0001-8384-5049}}
\affiliation{Department of Physics and Astronomy, West Virginia University, P.O. Box 6315, Morgantown, WV 26506, USA}
\affiliation{Center for Gravitational Waves and Cosmology, West Virginia University, Chestnut Ridge Research Building, Morgantown, WV 26505, USA}
\author{Gabriel E. Freedman\,\orcidlink{0000-0001-7624-4616}}
\affiliation{Center for Gravitation, Cosmology and Astrophysics, Department of Physics, University of Wisconsin-Milwaukee,\\ P.O. Box 413, Milwaukee, WI 53201, USA}
\author{Nate Garver-Daniels\,\orcidlink{0000-0001-6166-9646}}
\affiliation{Department of Physics and Astronomy, West Virginia University, P.O. Box 6315, Morgantown, WV 26506, USA}
\affiliation{Center for Gravitational Waves and Cosmology, West Virginia University, Chestnut Ridge Research Building, Morgantown, WV 26505, USA}
\author{Peter A. Gentile\,\orcidlink{0000-0001-8158-683X}}
\affiliation{Department of Physics and Astronomy, West Virginia University, P.O. Box 6315, Morgantown, WV 26506, USA}
\affiliation{Center for Gravitational Waves and Cosmology, West Virginia University, Chestnut Ridge Research Building, Morgantown, WV 26505, USA}
\author{Joseph Glaser\,\orcidlink{0000-0003-4090-9780}}
\affiliation{Department of Physics and Astronomy, West Virginia University, P.O. Box 6315, Morgantown, WV 26506, USA}
\affiliation{Center for Gravitational Waves and Cosmology, West Virginia University, Chestnut Ridge Research Building, Morgantown, WV 26505, USA}
\author{Deborah C. Good\,\orcidlink{0000-0003-1884-348X}}
\affiliation{Department of Physics, University of Connecticut, 196 Auditorium Road, U-3046, Storrs, CT 06269-3046, USA}
\affiliation{Center for Computational Astrophysics, Flatiron Institute, 162 5th Avenue, New York, NY 10010, USA}
\author{Kayhan G\"{u}ltekin\,\orcidlink{0000-0002-1146-0198}}
\affiliation{Department of Astronomy and Astrophysics, University of Michigan, Ann Arbor, MI 48109, USA}
\author{Ross J. Jennings\,\orcidlink{0000-0003-1082-2342}}
\altaffiliation{NANOGrav Physics Frontiers Center Postdoctoral Fellow}
\affiliation{Department of Physics and Astronomy, West Virginia University, P.O. Box 6315, Morgantown, WV 26506, USA}
\affiliation{Center for Gravitational Waves and Cosmology, West Virginia University, Chestnut Ridge Research Building, Morgantown, WV 26505, USA}
\author{Megan L. Jones\,\orcidlink{0000-0001-6607-3710}}
\affiliation{Center for Gravitation, Cosmology and Astrophysics, Department of Physics, University of Wisconsin-Milwaukee,\\ P.O. Box 413, Milwaukee, WI 53201, USA}
\author{Andrew R. Kaiser\,\orcidlink{0000-0002-3654-980X}}
\affiliation{Department of Physics and Astronomy, West Virginia University, P.O. Box 6315, Morgantown, WV 26506, USA}
\affiliation{Center for Gravitational Waves and Cosmology, West Virginia University, Chestnut Ridge Research Building, Morgantown, WV 26505, USA}
\author{David L. Kaplan\,\orcidlink{0000-0001-6295-2881}}
\affiliation{Center for Gravitation, Cosmology and Astrophysics, Department of Physics, University of Wisconsin-Milwaukee,\\ P.O. Box 413, Milwaukee, WI 53201, USA}
\author{Luke Zoltan Kelley\,\orcidlink{0000-0002-6625-6450}}
\affiliation{Department of Astronomy, University of California, Berkeley, 501 Campbell Hall \#3411, Berkeley, CA 94720, USA}
\author{Matthew Kerr\,\orcidlink{0000-0002-0893-4073}}
\affiliation{Space Science Division, Naval Research Laboratory, Washington, DC 20375-5352, USA}
\author{Joey S. Key\,\orcidlink{0000-0003-0123-7600}}
\affiliation{University of Washington Bothell, 18115 Campus Way NE, Bothell, WA 98011, USA}
\author{Nima Laal\,\orcidlink{0000-0002-9197-7604}}
\affiliation{Department of Physics, Oregon State University, Corvallis, OR 97331, USA}
\author{Michael T. Lam\,\orcidlink{0000-0003-0721-651X}}
\affiliation{SETI Institute, 339 N Bernardo Ave Suite 200, Mountain View, CA 94043, USA}
\affiliation{School of Physics and Astronomy, Rochester Institute of Technology, Rochester, NY 14623, USA}
\affiliation{Laboratory for Multiwavelength Astrophysics, Rochester Institute of Technology, Rochester, NY 14623, USA}
\author{William G. Lamb\,\orcidlink{0000-0003-1096-4156}}
\affiliation{Department of Physics and Astronomy, Vanderbilt University, 2301 Vanderbilt Place, Nashville, TN 37235, USA}
\author{T. Joseph W. Lazio\,\orcidlink}
\affiliation{Jet Propulsion Laboratory, California Institute of Technology, 4800 Oak Grove Drive, Pasadena, CA 91109, USA}
\author{Natalia Lewandowska\,\orcidlink{0000-0003-0771-6581}}
\affiliation{Department of Physics, State University of New York at Oswego, Oswego, NY, 13126, USA}
\author{Tingting Liu\,\orcidlink{0000-0001-5766-4287}}
\affiliation{Department of Physics and Astronomy, West Virginia University, P.O. Box 6315, Morgantown, WV 26506, USA}
\affiliation{Center for Gravitational Waves and Cosmology, West Virginia University, Chestnut Ridge Research Building, Morgantown, WV 26505, USA}
\author{Duncan R. Lorimer\,\orcidlink{0000-0003-1301-966X}}
\affiliation{Department of Physics and Astronomy, West Virginia University, P.O. Box 6315, Morgantown, WV 26506, USA}
\affiliation{Center for Gravitational Waves and Cosmology, West Virginia University, Chestnut Ridge Research Building, Morgantown, WV 26505, USA}
\author{Jing Luo\,\orcidlink{0000-0001-5373-5914}}
\altaffiliation{Deceased}
\affiliation{Department of Astronomy \& Astrophysics, University of Toronto, 50 Saint George Street, Toronto, ON M5S 3H4, Canada}
\author{Ryan S. Lynch\,\orcidlink{0000-0001-5229-7430}}
\affiliation{Green Bank Observatory, P.O. Box 2, Green Bank, WV 24944, USA}
\author{Chung-Pei Ma\,\orcidlink{0000-0002-4430-102X}}
\affiliation{Department of Astronomy, University of California, Berkeley, 501 Campbell Hall \#3411, Berkeley, CA 94720, USA}
\affiliation{Department of Physics, University of California, Berkeley, CA 94720, USA}
\author{Dustin R. Madison\,\orcidlink{0000-0003-2285-0404}}
\affiliation{Department of Physics, University of the Pacific, 3601 Pacific Avenue, Stockton, CA 95211, USA}
\author{Alexander McEwen\,\orcidlink{0000-0001-5481-7559}}
\affiliation{Center for Gravitation, Cosmology and Astrophysics, Department of Physics, University of Wisconsin-Milwaukee,\\ P.O. Box 413, Milwaukee, WI 53201, USA}
\author{James W. McKee\,\orcidlink{0000-0002-2885-8485}}
\affiliation{E.A. Milne Centre for Astrophysics, University of Hull, Cottingham Road, Kingston-upon-Hull, HU6 7RX, UK}
\affiliation{Centre of Excellence for Data Science, Artificial Intelligence and Modelling (DAIM), University of Hull, Cottingham Road, Kingston-upon-Hull, HU6 7RX, UK}
\author{Maura A. McLaughlin\,\orcidlink{0000-0001-7697-7422}}
\affiliation{Department of Physics and Astronomy, West Virginia University, P.O. Box 6315, Morgantown, WV 26506, USA}
\affiliation{Center for Gravitational Waves and Cosmology, West Virginia University, Chestnut Ridge Research Building, Morgantown, WV 26505, USA}
\author{Natasha McMann\,\orcidlink{0000-0002-4642-1260}}
\affiliation{Department of Physics and Astronomy, Vanderbilt University, 2301 Vanderbilt Place, Nashville, TN 37235, USA}
\author{Bradley W. Meyers\,\orcidlink{0000-0001-8845-1225}}
\affiliation{Department of Physics and Astronomy, University of British Columbia, 6224 Agricultural Road, Vancouver, BC V6T 1Z1, Canada}
\affiliation{International Centre for Radio Astronomy Research, Curtin University, Bentley, WA 6102, Australia}

\author{Chiara M. F. Mingarelli\,\orcidlink{0000-0002-4307-1322}}
\affiliation{Center for Computational Astrophysics, Flatiron Institute, 162 5th Avenue, New York, NY 10010, USA}
\affiliation{Department of Physics, University of Connecticut, 196 Auditorium Road, U-3046, Storrs, CT 06269-3046, USA}
\affiliation{Department of Physics, Yale University, New Haven, CT 06520, USA}
\author{Andrea Mitridate\,\orcidlink{0000-0003-2898-5844}}
\affiliation{Deutsches Elektronen-Synchrotron DESY, Notkestr. 85, 22607 Hamburg, Germany}
\author{Cherry Ng\,\orcidlink{0000-0002-3616-5160}}
\affiliation{Dunlap Institute for Astronomy and Astrophysics, University of Toronto, 50 St. George St., Toronto, ON M5S 3H4, Canada}
\author{David J. Nice\,\orcidlink{0000-0002-6709-2566}}
\affiliation{Department of Physics, Lafayette College, Easton, PA 18042, USA}
\author{Stella Koch Ocker\,\orcidlink{0000-0002-4941-5333}}
\affiliation{Cornell Center for Astrophysics and Planetary Science and Department of Astronomy, Cornell University, Ithaca, NY 14853, USA}

\author{Timothy T. Pennucci\,\orcidlink{0000-0001-5465-2889}}
\affiliation{Institute of Physics and Astronomy, E\"{o}tv\"{o}s Lor\'{a}nd University, P\'{a}zm\'{a}ny P. s. 1/A, 1117 Budapest, Hungary}
\author{Benetge B. P. Perera\,\orcidlink{0000-0002-8509-5947}}
\affiliation{Arecibo Observatory, HC3 Box 53995, Arecibo, PR 00612, USA}
\author{Nihan S. Pol\,\orcidlink{0000-0002-8826-1285}}
\affiliation{Department of Physics and Astronomy, Vanderbilt University, 2301 Vanderbilt Place, Nashville, TN 37235, USA}
\author{Henri A. Radovan\,\orcidlink{0000-0002-2074-4360}}
\affiliation{Department of Physics, University of Puerto Rico, Mayag\"{u}ez, PR 00681, USA}
\author{Scott M. Ransom\,\orcidlink{0000-0001-5799-9714}}
\affiliation{National Radio Astronomy Observatory, 520 Edgemont Road, Charlottesville, VA 22903, USA}
\author{Paul S. Ray\,\orcidlink{0000-0002-5297-5278}}
\affiliation{Space Science Division, Naval Research Laboratory, Washington, DC 20375-5352, USA}

\author{Shashwat C. Sardesai\,\orcidlink{0009-0006-5476-3603}}
\affiliation{Center for Gravitation, Cosmology and Astrophysics, Department of Physics, University of Wisconsin-Milwaukee,\\ P.O. Box 413, Milwaukee, WI 53201, USA}
\author{Carl Schmiedekamp\,\orcidlink{0000-0002-1283-2184}}
\affiliation{Department of Physics, Penn State Abington, Abington, PA 19001, USA}
\author{Ann Schmiedekamp\,\orcidlink{0000-0003-4391-936X}}
\affiliation{Department of Physics, Penn State Abington, Abington, PA 19001, USA}
\author{Kai Schmitz\,\orcidlink{0000-0003-2807-6472}}
\affiliation{Institute for Theoretical Physics, University of M\"{u}nster, 48149 M\"{u}nster, Germany}
\author{Brent J. Shapiro-Albert\,\orcidlink{0000-0002-7283-1124}}
\affiliation{Department of Physics and Astronomy, West Virginia University, P.O. Box 6315, Morgantown, WV 26506, USA}
\affiliation{Center for Gravitational Waves and Cosmology, West Virginia University, Chestnut Ridge Research Building, Morgantown, WV 26505, USA}
\affiliation{Giant Army, 915A 17th Ave, Seattle WA 98122}

\author{Joseph Simon\,\orcidlink{0000-0003-1407-6607}}
\altaffiliation{NSF Astronomy and Astrophysics Postdoctoral Fellow}
\affiliation{Department of Astrophysical and Planetary Sciences, University of Colorado, Boulder, CO 80309, USA}
\author{Magdalena S. Siwek\,\orcidlink{0000-0002-1530-9778}}
\affiliation{Center for Astrophysics, Harvard University, 60 Garden St, Cambridge, MA 02138}
\author{Ingrid H. Stairs\,\orcidlink{0000-0001-9784-8670}}
\affiliation{Department of Physics and Astronomy, University of British Columbia, 6224 Agricultural Road, Vancouver, BC V6T 1Z1, Canada}
\author{Daniel R. Stinebring\,\orcidlink{0000-0002-1797-3277}}
\affiliation{Department of Physics and Astronomy, Oberlin College, Oberlin, OH 44074, USA}
\author{Kevin Stovall\,\orcidlink{0000-0002-7261-594X}}
\affiliation{National Radio Astronomy Observatory, 1003 Lopezville Rd., Socorro, NM 87801, USA}
\author{Abhimanyu Susobhanan\,\orcidlink{0000-0002-2820-0931}}
\affiliation{Center for Gravitation, Cosmology and Astrophysics, Department of Physics, University of Wisconsin-Milwaukee,\\ P.O. Box 413, Milwaukee, WI 53201, USA}
\author{Joseph K. Swiggum\,\orcidlink{0000-0002-1075-3837}}
\altaffiliation{NANOGrav Physics Frontiers Center Postdoctoral Fellow}
\affiliation{Department of Physics, Lafayette College, Easton, PA 18042, USA}
\author{Jacob E. Turner\,\orcidlink{0000-0002-2451-7288}}
\affiliation{Department of Physics and Astronomy, West Virginia University, P.O. Box 6315, Morgantown, WV 26506, USA}
\affiliation{Center for Gravitational Waves and Cosmology, West Virginia University, Chestnut Ridge Research Building, Morgantown, WV 26505, USA}
\author{Caner Unal\,\orcidlink{0000-0001-8800-0192}}
\affiliation{Department of Physics, Ben-Gurion University of the Negev, Be'er Sheva 84105, Israel}
\affiliation{Feza Gursey Institute, Bogazici University, Kandilli, 34684, Istanbul, Turkey}
\author{Haley M. Wahl\,\orcidlink{0000-0001-9678-0299}}
\affiliation{Department of Physics and Astronomy, West Virginia University, P.O. Box 6315, Morgantown, WV 26506, USA}
\affiliation{Center for Gravitational Waves and Cosmology, West Virginia University, Chestnut Ridge Research Building, Morgantown, WV 26505, USA}
\author{Caitlin A. Witt\,\orcidlink{0000-0002-6020-9274}}
\affiliation{Center for Interdisciplinary Exploration and Research in Astrophysics (CIERA), Northwestern University, Evanston, IL 60208}
\affiliation{Adler Planetarium, 1300 S. DuSable Lake Shore Dr., Chicago, IL 60605, USA}
\author{Olivia Young\,\orcidlink{0000-0002-0883-0688}}
\affiliation{School of Physics and Astronomy, Rochester Institute of Technology, Rochester, NY 14623, USA}
\affiliation{Laboratory for Multiwavelength Astrophysics, Rochester Institute of Technology, Rochester, NY 14623, USA}
\collaboration{NANOGrav Collaboration}

\date{\today}

\begin{abstract}
Pulsar timing arrays (PTAs) use an array of millisecond pulsars to search for gravitational waves in the nanohertz regime in pulse time of arrival data.
This paper presents rigorous tests of PTA methods, examining their consistency across the relevant parameter space.
We discuss updates to the 15-year isotropic gravitational-wave background analyses and their corresponding code representations.
Descriptions of the internal structure of the flagship algorithms \texttt{Enterprise} and \texttt{PTMCMCSampler} are given to facilitate understanding of the PTA likelihood structure, how models are built, and what methods are currently used in sampling the high-dimensional PTA parameter space.
We introduce a novel version of the PTA likelihood that uses a two-step marginalization procedure that performs much faster in gravitational wave searches, reducing the required resources facilitating the computation of Bayes factors via thermodynamic integration and sampling a large number of realizations for computing Bayesian false-alarm probabilities.
We perform stringent tests of consistency and correctness of the Bayesian and frequentist analysis methods.
For the Bayesian analysis, we test prior recovery, simulation recovery, and Bayes factors.
For the frequentist analysis, we test that the optimal statistic, when modified to account for a non-negligible gravitational-wave background, accurately recovers the amplitude of the background.
We also summarize recent advances and tests performed on the optimal statistic in the literature from both GWB detection and parameter estimation perspectives.
The tests presented here validate current analyses of PTA data.
\end{abstract}
\maketitle


\section{Introduction}
\label{sec:intro}
Since the first detection of gravitational waves (GWs) from a stellar mass black hole binary in 2015~\cite{abbott_observation_2016-1}, the field of GW astronomy has flourished with dozens more detections of transient signals~\cite{the_ligo_scientific_collaboration_gwtc-3_2021-1}.
Besides these signals at ${\cal{O}}(100)$\,Hz frequencies, detected via ground-based laser interferometry, other methods can detect GWs across a wide range of frequencies.
Pulsar timing arrays (PTAs) create a galactic-scale GW detector using radio telescopes to collect pulse times of arrival (TOAs) from an array of millisecond pulsars.
These pulsars exhibit exceptional long-timescale arrival-time stability, allowing PTAs to use them as galactic scale clocks that are sensitive to perturbations at frequencies $1$--$100$\,nHz.
The expected target signal is the stochastic GW background (GWB), possibly from an ensemble of merging supermassive black-hole binaries that could result from galactic mergers~\cite{burke-spolaor_astrophysics_2019}.

Previously, the European Pulsar Timing Array (EPTA)~\cite{10.1093/mnras/stw483}, the Parkes Pulsar Timing Array in Australia (PPTA)~\cite{manchester_parkes_2013}, and the North American Nanohertz Observatory for Gravitational waves (NANOGrav)~\cite{mclaughlin_north_2013,brazier2019nanograv} all reported detection of a red-noise process with common spectrum among pulsars, but no evidence either way for inter-pulsar correlations~\cite{NG12.gwb, chen_common-red-signal_2021, goncharov_evidence_2021}.
Such a process is known as CURN, for Common-spectrum Uncorrelated Red Noise.
The International Pulsar Timing Array (IPTA)~\cite{perera_international_2019} consists of these collaborations along with the Indian Pulsar Timing Array (InPTA)~\cite{joshi_precision_2018} and the MeerKAT Pulsar Timing Array (MPTA)~\cite{spiewak_meertime_2022}.
Combining data from older data sets from NANOGrav, the EPTA, and the PPTA, the IPTA also found a consistent CURN in their second data release~\cite{antoniadis_international_2022}.
Such a signal might arise from some currently unknown noise processes shared by the measured pulsars~\cite{goncharov_evidence_2021, zic_evaluating_2022, goncharov_consistency_2022}.
Thus, to claim a detection of GWs we require evidence of the telltale correlations between pulsar pairs, known as Hellings and Downs (HD) correlations~\cite{hellings_upper_1983}.

Among the 67 pulsars used in the GWB analysis in the 15-year data~\cite{NG15.data}, NANOGrav reports a stronger detection of the CURN process.
Additionally, NANOGrav reports evidence for HD correlations suggesting that the common signal seen by the three PTAs is indeed a gravitational-wave background~\cite{NG15.GWB}.
While the parameter estimation and subsequent astrophysical interpretation remains somewhat dependent on noise models (see, e.g.,~\cite{NG15.detchar} for more information about the noise models used), the GWB remains consistent with an origin of an ensemble of supermassive black hole binaries~\cite{NG15.astro}.
In addition to searching for a GWB, other analyses have been performed looking for single continuous-wave sources~\cite{NG15.contwave}, anisotropy in the GWB~\cite{NG15.anisotropy}, and possible hints of new physics~\cite{NG15.NP}.
Future joint analyses in the IPTA will work toward combining data to improve sensitivity to GW sources and their corresponding parameters for astrophysical interpretation.

In performing analyses for HD correlations, PTA collaborations employ a variety of techniques to process and analyze their data.
NANOGrav uses a modular pipeline that starts with radio telescope data and ends with astrophysical GWB inference.
Radio telescope data are processed into TOAs, analyzed for outliers, and fitted with a timing model \cite{NG15.data}.
Subtracting the predicted TOAs from the observed TOAs results in the TOA \emph{residuals}.
These residuals are broken down into terms associated with deterministic signals, as one might expect from an individual binary, as well as stochastic signals such as those from a GWB.
Stochastic signals may be further broken into different types of noise processes: white and red.
White noise has a flat power spectral density and is associated with noise in telescope observations.
Red noise has a larger amplitude at lower frequencies and originates from pulsars themselves in the form of spin noise \cite{spin_noise}, fluctuations in the dispersion measure (DM) caused by the interstellar medium in between us and the pulsar, and a potential GWB \cite{dm_variations}.
A single-pulsar white and red noise analysis is performed at the end of this part in the pipeline.
The TOAs, timing model, residuals, and noise analyses comprise the initial input for any GW analysis.

Unlike other regimes of GW data analysis, most current PTA gravitational-wave analyses are performed in the time domain~\cite{chamberlin_time-domain_2015,van_haasteren_measuring_2009}.
In this paper, we specialize to the case of a GWB and describe the analysis implementations as they currently exist.
The Bayesian GWB analysis started with the pioneering work of van Haasteren et. al~\cite{van_haasteren_measuring_2009}.
While the brute force inversion of a full $N_\text{TOA} \times N_\text{TOA}$ matrix 
was possible with $\sim 10^3$ TOAs at the time of publication, it would not work today with $\sim 5 \times 10^5$ TOAs.
The modern PTA likelihood was introduced in~\cite{Lentati+2013} where the authors expanded the red noise in a set of Fourier coefficients.
\citet{Lentati+2013} marginalized over the timing model as in~\cite{van_haasteren_measuring_2009} and added an additional marginalization over the Fourier coefficients.
\citet{vanHaasVallis2014, vanHaasVallis2015} used the connection between these methods and the theory of Gaussian processes leading to further optimizations in the likelihood computation and sampling.

Complementary to Bayesian approaches, the NANOGrav frequentist GWB analysis uses the so-called optimal statistic (OS), an estimator of the amplitude and significance of a GWB.
Initially, this statistic was formulated in the frequency domain~\cite{anholm_optimal_2009}, and later it was re-implemented in the time domain~\cite{chamberlin_time-domain_2015}. Traditionally, it has been formulated in the regime where the amplitude of the GWB is much lower than the amplitude of the intrinsic red noise in the pulsars, an assumption that has been relaxed recently~\cite{Allen:2022ksj}. Additionally, the optimal statistic relies upon an estimate of the intrinsic red noise in each pulsar. Our estimates of intrinsic red noise are uncertain, and using a specific choice of the red noise can result in a biased statistic~\cite{vigeland_noise-marginalized_2018}.
The solution is to create a hybrid Bayesian-frequentist ``noise-marginalized'' optimal statistic, in which the optimal statistic is computed over many posterior draws for our red noise model, creating a distribution for the amplitude estimate of the GWB and its associated significance. It is common practice to average the optimal statistic over this distribution~\cite{vigeland_noise-marginalized_2018}, although recently alternative approaches have also been proposed~\cite{ppc1}.

The traditional optimal statistic assumes that there is only one type of spatial correlation in the data.
Monopolar and dipolar correlations are also possible, resulting from systematic issues such as clock errors~\cite{tiburzi_study_2016} or solar system ephemeris errors~\cite{vallisneri_modeling_2020} respectively, and so potentially there will be multiple spatially correlated signals in the data simultaneously.
When searching for only one type of spatially correlated signal at a time, one might make spurious detections due to the contribution of other spatially correlated signals.
The solution to this is to simultaneously fit for multiple spatial correlation patterns using the multiple-component optimal statistic (MCOS)~\cite{sardesai_generalized_2023}.

These methods have evolved conceptually over the past decade, and so have their software implementations.
NANOGrav maintains a set of publicly available software packages that are written in the \texttt{Python} programming language and make extensive use of \texttt{numpy} and \texttt{scipy}~\cite{numpy, scipy}.
The packages that we focus on in this work include \texttt{Enterprise} \github{https://github.com/nanograv/enterprise}, \texttt{enterprise\_extensions} \github{https://github.com/nanograv/enterprise_extensions}, \texttt{PTMCMCSampler} \github{https://github.com/jellis18/PTMCMCSampler}, and \texttt{la\_forge} \github{https://github.com/nanograv/la_forge}~\cite{enterprise, enterprise-extensions, ptmcmcsampler, hazboun_forge_2020}.
In this paper, we review these packages and perform a campaign of simulations to validate them for the specific case of searching for a GWB with NANOGrav data, as performed in NANOGrav's 15-year GWB analysis \cite{NG15.GWB}.
We base all simulations in this paper on the NANOGrav 15-year data set, using all 67 pulsars that have been timed for more than three years \cite{NG15.data}.

Inspired by probabilistic programming packages such as Stan~\cite{lee_stan-devstan_2017} and PyMC~\cite{wiecki_pymc-devspymc_2023}, \texttt{Enterprise} allows the specification of probabilistic data models for PTAs, and the evaluation of the resulting priors and likelihoods.
By contrast, \texttt{enterprise\_extensions} contains pre-built models that are commonly used in PTA analyses.
It also includes \emph{hypermodels} used in Bayesian model selection by way of product-space sampling, as well as implementations of the OS, MCOS, and a noise-averaged version of these known as the ``noise-marginalized'' optimal statistic (NMOS).
We use \texttt{PTMCMCSampler}, a parallel-tempering enabled Markov-chain Monte Carlo (MCMC) sampler, to approximate the posterior of the models created with \texttt{Enterprise}.
Finally, once we finish sampling, we use \texttt{la\_forge} to compress and post-process the resulting chains.

We address two main issues in this work.
First, as the number of pulsars in our data set increases, our array gains sensitivity and the computations become longer due to an increased number of parameters and the increased cost of likelihood calculation.
Robust Bayes factor calculations such as thermodynamic integration also require large amounts of computational resources which were previously prohibitive, taking months to complete before speeding up the likelihood calculation.
Bayes factors between many of the models computed in the 15-year analysis provide an intractable challenge for the standard product-space model comparisons which have been used in previous analyses due to the large Bayes factors and difficulty of finding a good weight to allow for the chains to switch between the models easily.
Thermodynamic integration solves this issue and is especially necessary in the cases where the Bayes factor is very large or very small.
Performing a large series of Bayesian analyses, such as are used in the Bayesian false-alarm probability calculations~\cite{NG15.GWB}, is also computationally prohibitive.
To speed up these computations, we implement a $\sim 5\times$ faster method of likelihood computation, which is equivalent mathematically to the previous method.
Second, the original optimal statistic for PTAs~\cite{chamberlin_time-domain_2015} works well as a detection statistic under the no-signal null hypothesis, but remains biased as an estimator for the GWB amplitude.
We check that this bias has been reduced by accounting for a GWB.
Both of these methods provide crucial adjustments to the NANOGrav methods for future analyses, as the number of pulsars increases and the background becomes even more prominent in our dataset.

In \autoref{sec:enterprise}, we describe the traditional likelihood computation and a new faster implementation for situations where the white-noise parameters remain constant. We also describe the computational scaling associated with each calculation.
In \autoref{sec:Bayes} we discuss using \texttt{Enterprise} along with \texttt{PTMCMCSampler} to explore the high dimensional parameter spaces in GWB analyses.
Frequentist methods implemented in \texttt{enterprise\_extensions} are discussed in \autoref{sec:frequentist}.
Next, we present tests on the Bayesian methods in \autoref{sec:bayes_tests} and frequentist methods in \autoref{sec:frequentist_tests}.
Finally, we discuss possible future directions and conclude in \autoref{sec:concl}.

\section{The PTA Likelihood Calculations}
\label{sec:enterprise}
First, we describe the standard and ``fast'' PTA likelihoods.
Both of these likelihoods use \texttt{Enterprise}, a pure \texttt{Python} package built to analyze pulsar noise, timing models, and to search for GWs in PTA data.
For a discussion of \texttt{Enterprise} and its structure, see App.~\ref{app:struct_enterprise}.

\subsection{The PTA Likelihood}
\label{sec:likelihood}

Our TOAs $\vect{t}$ can be written as
\begin{equation}
    \vect{t} = \vect{t}_{\text{det}} + \vect{t}_{\text{stoc}}\,,
\end{equation}
where $\vect{t}_{\text{det}}$ is the deterministic part of the TOAs, $\vect{t}_{\text{stoc}}$ is the stochastic part of the TOAs.
After fitting the deterministic part of the signal with a least squares fit, $\vect{t}_\text{M}$,
\begin{equation}
    \vect{t}_\text{det} \approx \vect{t}_\text{M} + \vect{M}\vect{\epsilon}\,,
\end{equation}
where $\vect{M}\vect{\epsilon}$ is a Taylor expansion of the residual deterministic part where $\vect{M}$ consists of derivatives in the Taylor expansion.
The details of the process used to produce the timing model in the NANOGrav 15-year data set can be found in the data set paper \cite{NG15.data}.
The stochastic part of the TOAs
\begin{equation}
    \vect{t}_\text{stoc} = \vect{F c} + \vect{n}\,,
\end{equation}
where $\vect{c}$ represent the Fourier coefficients, $\vect{F}$ represents a discrete Fourier transform of the red noise processes in the data, and $\vect{n}$ consists of white noise.
Combining all of the above and subtracting off the timing model fit, we find
\begin{equation}
    \vect{\delta t} = \vect{t} - \vect{t}_\text{M} \approx \vect{M \epsilon} + \vect{F c} + \vect{n}\,.
\end{equation}
As in~\cite{Lentati+2013}, we expand red noise in a set of Fourier coefficients that determine a specific random realization of a stochastic process,
\begin{equation}
    \vect{F c}=\sum_{j=1}^N\left[X_j \sin \left(2 \pi f_j t\right)+Y_j \cos \left(2 \pi f_j t\right)\right]\,,
\end{equation}
where alternating $\vect{X}, \vect{Y}$ make up $\vect{c}$, $\vect{F}$ contains alternating columns of sine and cosine components, and $f_i = i / T$ with $T$ the observing time span of the entire data set (16.03 years in the 15-year data set \footnote{The data set has been so-named, because the pulsar with the longest observation time span contains 15.8 years of data.}).
Red noise may consist of components intrinsic to the pulsar, such as spin noise, or even a gravitational-wave background.
We limit the number of frequency bins $N$ used based on the model of each red noise signal.
For pulsar intrinsic red noise, we limit ourselves to 30 frequency bins which is sufficient to capture the high-frequency content of the data.
This corresponds to frequencies from about 2 nHz to about 60 nHz.
The number of frequencies used in the GWB analyses depends on how we model it.
For example, a two-parameter power law with amplitude and spectral index in the 15-year data set shows that the red noise dips below the white noise at around 14 frequency bins or around 28 nHz \cite{NG15.GWB}.
While we could include more frequency bins (50 were included in~\cite{NG11.gwb}), adding more frequencies on this model would bias the common power law spectral index and amplitude, unless a more advanced red noise model is used \cite{adv_noise12.5}.
Using a model that allows each frequency bin to vary independently is known as a ``free spectrum model''.
For this model, bins are not affected significantly by adjacent bins, and we use the same 30 frequency bins on the GWB that we use on the intrinsic red noise.
Therefore, including more frequency bins on common signals has negligible influence on a free spectrum analysis, but this model also includes an additional parameter per frequency bin.

Several types of white noise parameters are searched over that adjust the TOA uncertainties found during template fitting, e.g.,~\cite{ng11_data}.
EQUAD $\vect{Q}$ is an extra factor added in quadrature to the TOA uncertainties.
EFAC $\vect{G}$ rescales the TOA uncertainties and EQUAD together.
Finally, ECORR $\vect{J}$ is an extra term that correlates different frequency bands within the same epoch, a term which here means a single observation.
A single observation (epoch) could be a single day, or it could be a combination of multi-band data from separate days combined.
Separate epochs remain completely uncorrelated.
ECORR can account for pulse jitter \cite{NG15.detchar}.
Each of these white noise signals add one dimension per pulsar per observing backend per frequency band used to acquire data.
The white noise covariance matrix can be written as
\begin{equation}
    \vect{N} = \left<\vect{n} \vect{n}^T\right> = \sum_{\mu}\left[ G_{\mu}^2\left(\sigma_i^2 + Q_{\mu}^2\right)\delta_{ij} + J_{\mu}^2\delta_{e(i)e(j)}\right]\,,
    \label{eq:whitenoise}
\end{equation}
where the $i$th TOA belongs to the backend $\mu$, $\delta_{ij}$ denotes the Kronecker delta, and $\delta_{e(i)e(j)}$ denotes another Kronecker delta that equals 1 only when the epochs are the same for both TOAs considered and 0 otherwise.

Subtracting the deterministic and stochastic models results in the residual,
\begin{equation}
    \vect{r} = \vect{\delta t} - \vect{M\epsilon} - \vect{F c}\,.
\end{equation}
Under a multi-variate Gaussian assumption for the noise the full likelihood can then be written as
\begin{equation}
    \label{eq:start_like}
    p(\vect{\delta t} \mid \vect{c}, \vect{\epsilon})=\frac{1}{\sqrt{\det(2 \pi \vect{N})}} \exp\left({-\frac{1}{2}\vect{r}^T \vect{N}^{-1}\vect{r}}\right)\,.
\end{equation}
We then group the matrices and vectors into a more compact notation,
\begin{align}
    \vect{T} = \begin{bmatrix}
    \vect{M} & \vect{F}
    \end{bmatrix}\,,
    &&
    \vect{b} = \begin{bmatrix}
    \vect{\epsilon}\\
    \vect{c}
\end{bmatrix}\,,
\end{align}
and place a Gaussian prior on the Fourier coefficients with covariance
\begin{equation}
    \vect{B}(\vect{\eta}) = \begin{bmatrix}
        \vect{E} & 0 \\
        0 & \vect{\phi}(\vect{\eta})
    \end{bmatrix} = \begin{bmatrix}
        \vect{\infty} & 0 \\
        0 & \vect{\phi}(\vect{\eta})
    \end{bmatrix}\,,
\end{equation}
where $\vect{\eta}$ contains the hyper-parameters, i.e., the parameters of the prior such as amplitudes or spectral indices of the GWB or red noise power laws. 
We let $\vect{E}$ be a diagonal matrix of very large values ($10^{40}$ by default) in $\vect{B}$.
This places an improper, almost-infinite variance Gaussian prior on the timing model parameters.
However, these parameters are well determined by pulsar timing observations and thus likelihood-dominated.
Upon inversion, this choice marginalizes over the timing model uncertainties~\cite{Lentati+2013, vanHaasVallis2014, vanHaasVallis2015}.

The covariance matrix for the red noise coefficients,
\begin{equation}
    \vect{\phi} = \left<\vect{c}\vect{c}^T\right>\,,
\end{equation}
can be constructed via blocks.
Each block contains
\begin{equation}
    N_{\text{freq}} = \max\left(N_{\text{IRN}}, N_{\text{GWB}}\right)\,,
\end{equation}
frequencies where $N_{\text{IRN}}$ is the number of frequencies used on the intrinsic red noise, and $N_{\text{GWB}}$ is the number of frequencies used on a red noise process common among pulsars (correlated or not).
If we denote the block containing the frequencies of pulsars as $(a, b)$, and we further specify each frequency as $(i, j)$, then
\begin{equation}
    \left[\phi\right]_{(ai)(bj)}=\left\langle c_{a i} c_{b j}\right\rangle=\delta_{i j}\left(\delta_{a b} \varphi_{a i}+\Phi_{a b, i}\right),
\end{equation}
with
\begin{equation}
    \Phi_{a b, i} = \Gamma_{a b}\Phi_i,
\end{equation}
where the overlap reduction function (ORF), $\Gamma_{a b}$, describes the correlations between pulsar pairs.
In the isotropic GWB analysis, the ORFs we can search for include a no-correlation ORF for CURN, monopolar correlations which could be caused by clock corrections \cite{tiburzi_study_2016}, dipolar correlations which could be caused by solar system ephemeris errors \cite{vallisneri_modeling_2020}, and the HD correlations which are characteristic of a GWB:
\begin{align}
    \Gamma_{a b}^\text{CURN} &= \delta_{a b}\,,\\
    \Gamma_{a b}^\text{MON} &= 1\,,\\
    \Gamma_{a b}^\text{DIP} &= \cos{\theta_{ab}}\,,\\
    \Gamma_{a b}^\text{HD} &= \frac{1}{2}\delta_{a b} + \frac{3}{2}x_{ab} \ln x_{ab} - \frac{1}{4}x_{ab} + \frac{1}{2}\,,
\end{align}
where $x_{ab} = \left( 1 - \cos{\xi_{ab}} \right) / 2$ and $\xi_{ab}$ is the angle between pulsars $a$ and $b$ on the sky.
The terms $\rho$ and $\kappa_{a}$ are related to the power spectral density, $S(f)$ of the time delay caused by intrinsic red noise or a GWB respectively as
\begin{equation}
    \varphi_{ai}, \Phi_i = S(f_i)\Delta f = S(f) / T\,.
\end{equation}
A typical model choice for these spectra is a power law,
\begin{equation}
    \varphi_{ai}, \Phi_i = \frac{A^2}{12\pi^2}\frac{1}{T}\left(\frac{f_i}{f_{\text{ref}}}\right)^{-\gamma}\text{yr}^2\,,
    \label{eq:rho_kappa_definition}
\end{equation}
where the GWB characteristic strain is,
\begin{equation}
    h_c = A\left(\frac{f}{f_{\text{ref}}}\right)^\alpha\,,
\end{equation}
with $\gamma = 3 - 2\alpha$.
Provided that the GWB is made up of signals from an ensemble of supermassive black hole binaries, we expect $\gamma = 13/3$ ($\alpha = -2/3$) \cite{phinney_practical_2001}.
The reference frequency, $f_{\text{ref}}$, traditionally has been set to $1 / \text{yr}^{-1}$, and this is the value that we use in our simulations.
For the intrinsic red noise and common uncorrelated red process, $A$ and $\gamma$,
$A$ and $\gamma$ are the hyperparameters, $\vect{\eta}$, for a power law spectrum model.

In the traditional form of the likelihood, we analytically marginalize over $\vect{b}$.
This is a simultaneous marginalization over the timing model uncertainty and red noise coefficients,
\begin{equation}
    p(\vect{\delta t}\mid \vect{\eta}) = \int p(\vect{\delta t}\mid \vect{b}) p(\vect{b} \mid \vect{\eta})d\vect{b}\,,
\end{equation}
where
\begin{equation}
    p(\vect{b} \mid \vect{\eta}) = \frac{\exp\left(-\frac{1}{2}\vect{b}^T\vect{B}^{-1}\vect{b}\right)}{\sqrt{\det(2\pi \vect{B})}}\,,
\end{equation}
is a Gaussian prior with hyper-parameters $\vect{\eta}$.
Evaluating the integral gives
\begin{equation}
    \label{eq:trad_like}
    p(\vect{\delta t} \mid \vect{\eta})=\frac{1}{\sqrt{\det(2 \pi \vect{C})}}\exp \left(-\frac{1}{2} \vect{\delta t}^{\mathrm{T}} \vect{C}^{-1} \vect{\delta t}\right)\,,
\end{equation}
where the covariance matrix is now
\begin{equation}
    \vect{C} = \vect{N} + \vect{TBT}^T\,.
\end{equation}
Such an inversion can be efficiently evaluated with the Woodbury matrix identity,
\begin{equation}
    \vect{C}^{-1}=\vect{N}^{-1}-\vect{N}^{-1} \vect{T} \vect{\Sigma} \vect{T}^T \vect{N}^{-1}\,,
\end{equation}
with
\begin{equation}
    \vect{\Sigma} = \left(\vect{B}^{-1} + \vect{T}^T \vect{N}^{-1} \vect{T}\right)^{-1}\,.
\end{equation}
This formulation reduces the number of operations required for the computational bottleneck from an inversion of the full covariance matrix $\vect{C}$ which takes $\mathcal{O}\left(N_{\text{p}}^3 N_{\text{TOA}}^3\right)$ operations to the inversion of $\vect{\Sigma}$ which takes $\mathcal{O}\left(N_{\text{p}}^3(2 N_{\text{freq}} + N_\text{M})^3\right)$.
Given that the number of TOAs is $\sim 10^5$, $N_{\text{freq}} = 30$, $N_{\text{M}} \sim 10^2$, the savings are significant.

The white noise covariance matrix is block-diagonal with a block for each observation epoch.
We invert each block efficiently using the Sherman-Morrison formula
\begin{equation}
    \boldsymbol{N}_b^{-1}=\boldsymbol{H}_b^{-1}-\frac{\boldsymbol{H}_b^{-1} \vect{u} \vect{u}^{\mathrm{T}} \boldsymbol{H}_b^{-1}}{J^{-2}+\vect{u}^{\mathrm{T}} \boldsymbol{N}_b^{-1} \vect{u}}\,,
    \label{eq:shmo}
\end{equation}
where $\vect{H}_b$ contains the diagonal elements of the white noise covariance matrix, and
\begin{equation}
    \vect{u}^T = (1, 1, \cdots, 1),
\end{equation}
with length of the number of TOAs in the epoch.
While this could be a non-negligible part of the computation, we fix the white noise in the GWB analysis, and therefore we do not consider the speed of evaluation here.

\subsection{Faster Likelihood for GWB Analyses}
\label{sec:fasterlike}
White-noise parameters are varied in noise runs prior to performing any GWB analyses.
After these initial runs, all white noise parameters are set to their median marginalized posterior values to reduce the total number of parameters from $\sim 10^3$ to $\sim 10^2$. When modeling the stochastic processes as a power law, this results in 136 parameters: two parameters for each of the 67 pulsar's intrinsic red noise parameters, and two for the GWB.
White noise parameters in \texttt{Enterprise} are cached to speed up evaluation time significantly after the initial likelihood evaluation.

Starting with the same multi-variate Gaussian in \autoref{eq:start_like}, we now marginalize in two steps instead of simultaneously.
Marginalizing over the timing model,
\begin{equation}
    p(\vect{\delta t} \mid \vect{c}, \vect{E})=\int p(\vect{\delta t} \mid \vect{\epsilon}, \vect{c}) p(\vect{\epsilon} \mid \vect{\eta}) d \vect{\epsilon}\,,
\end{equation}
with the Gaussian prior
\begin{equation}
    p(\boldsymbol{\epsilon} \mid \boldsymbol{\eta})=\frac{\exp \left(-\frac{1}{2} \boldsymbol{\epsilon}^T \boldsymbol{E}^{-1} \boldsymbol{\epsilon}\right)}{\sqrt{\operatorname{det}(2 \pi \boldsymbol{E})}}\,,
\end{equation}
and $\vect{E} = \left<\vect{\epsilon}\vect{\epsilon}^T\right>$. This integral results in
\begin{equation}
    \label{eq:det_issue}
    p(\vect{\delta t} \mid \vect{c})=\frac{\exp{\left(-\frac{1}{2}(\vect{\delta t}-\vect{F} \vect{c})^T \vect{D}^{-1}(\vect{\delta t}-\vect{F} \vect{c})\right)}}{\sqrt{\det (2\pi \vect{D})} } \,,
\end{equation}
with
\begin{equation}
    \vect{D}=\vect{N}+\vect{M} \vect{E} \vect{M}^T\,.
\end{equation}
We use the Woodbury matrix identity for this inversion,
\begin{equation}
    \vect{D}^{-1}=\vect{N}^{-1}-\vect{N}^{-1} \vect{M} \vect{\Lambda}^{-1} \vect{M}^T \vect{N}^{-1}\,,
\label{eq:fastwoodbury}
\end{equation}
with
\begin{equation}
    \vect{\Lambda}=\vect{E}^{-1}+\vect{M}^T \vect{N}^{-1} \vect{M}=\vect{M}^T \vect{N}^{-1} \vect{M}\,,
\end{equation}
because $\diag(\vect{E}) \rightarrow \infty$\footnote{Formally, we use the matrix determinant lemma to evaluate the determinant in \autoref{eq:det_issue} as 
\begin{equation}
    \det(\vect{D}) = \det\left(\vect{E}^{-1} + \vect{M}^T\vect{N}^{-1}\vect{M}\right)\det(\vect{E})\det(\vect{N}),
\end{equation}
where $\diag(\vect{E}) = \infty$. However, in practice we can use $10^{40}$ as an effectively infinite value and avoid dealing with the infinite determinant term.}.
Now, we complete the two-step marginalization procedure by marginalizing over the Fourier coefficients,
\begin{equation}
    p(\vect{\delta t} \mid \vect{\eta}) = \int p(\vect{\delta t} \mid \vect{c}) p(\vect{c} \mid \vect{\eta}) d \vect{c}\,,
\end{equation}
with the Gaussian prior
\begin{equation}
    p(\boldsymbol{c} \mid \boldsymbol{\eta})=\frac{\exp \left(-\frac{1}{2} \boldsymbol{c}^T \boldsymbol{\phi}^{-1} \boldsymbol{c}\right)}{\sqrt{\operatorname{det}(2 \pi \boldsymbol{\phi})}}\,,
\end{equation}
resulting in
\begin{equation}
    \label{eq:fast_like}
    p(\vect{\delta t} \mid \vect{\eta}) = \frac{1}{\sqrt{\det(2 \pi \vect{K})}} \exp\left(-\frac{1}{2}\vect{\delta t}^T \vect{K}^{-1}\vect{\delta t}\right)\,,
\end{equation}
where
\begin{equation}
    \vect{K}=\vect{D}+\vect{F} \vect{\phi} \vect{F}^T,
\end{equation}
and once again we use the Woodbury matrix identity to invert this covariance matrix.
We find
\begin{equation}
    \vect{K}^{-1}=\vect{D}^{-1}-\vect{D}^{-1} \vect{F} \vect{\Theta} \vect{F}^T \vect{D}^{-1}\,,
\end{equation}
with
\begin{equation}
    \label{eq:phi_inv}
    \vect{\Theta}=\left(\vect{\phi}^{-1}+\vect{F}^T \vect{D}^{-1} \vect{F}\right)^{-1}\,.
\end{equation}
As long as $\vect{D}^{-1}$ can be cached in its entirety, the computation will be faster with the two-step marginalization, \autoref{eq:fast_like}, over the simultaneous marginalization procedure, \autoref{eq:trad_like}.
Now, the $\vect{\Theta}$ inversion in \autoref{eq:phi_inv} dominates the likelihood computation with $\mathcal{O}\left(N_p^3 (2 N_{\text{freq}})^3\right)$ operations.

\subsection{Empirical Computational Scaling for Likelihood Evaluations}
\begin{figure*}
    \centering
    \begin{subfigure}{\columnwidth}
        \centering
        \includegraphics[width=.97\columnwidth]{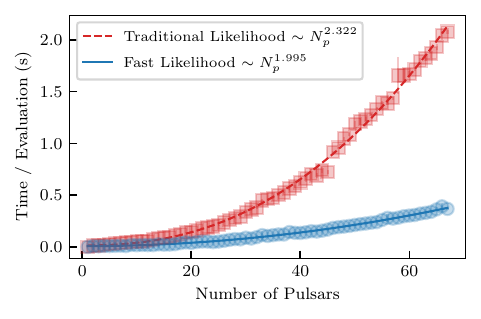}
        \caption{HD Model Comparison}
        \label{fig:corr}
    \end{subfigure}%
    \begin{subfigure}{\columnwidth}
        \centering
        \includegraphics[width=\columnwidth]{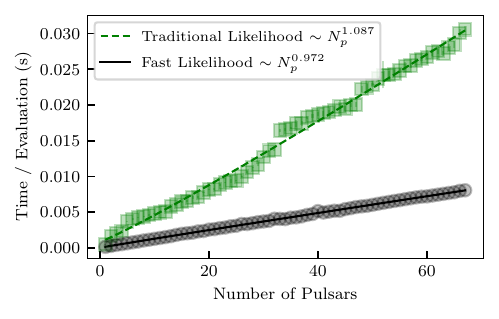}
        \caption{CURN Model Comparison}
        \label{fig:uncorr}
    \end{subfigure}%
    \caption{(a)~Comparison between the empirical scaling of HD-correlated models for both the simultaneous marginalization procedure (labeled ``traditional likelihood'') and the new two-step marginalization procedure (labeled ``fast likelihood'') which runs faster for models where the white noise is not being varied.
    In the latter case, we can cache the $\boldsymbol{D}$ matrix resulting in a drastic reduction in computation time at the cost of some memory.
    Inversions in both methods use a sparse Cholesky decomposition method on a single CPU core.
    This new marginalization procedure results in an average speed increase per evaluation of $5.65 \times$ for 67 pulsars.
    (b)~Comparison between the CURN models using the different marginalization procedures.
    Inverting the now diagonal matrix of these models is trivial.
    The reduction in number of elements required to be inverted results in a speed increase of $3.80 \times$ for 67 pulsars.
    Individual points and uncertainties are found by averaging over 100 evaluations of a model and taking the standard deviation.
    Fits to the points have been made with a nonlinear least squares algorithm fitting $A,B,C$ to a function $f(N_p) = A N_p^B + C$.
    }
    \label{fig:timing}
\end{figure*}

As shown in \autoref{fig:timing}, we find a significant improvement in the fast version of the likelihood, \autoref{eq:fast_like}, over the traditional likelihood \autoref{eq:trad_like}.
With 67 pulsars, in the HD-correlated model, the fast likelihood leads to 5.65 times faster evaluation than the traditional computation. The speed up factor is reduced to 3.8 in the CURN model.
Empirically, as the number of pulsars increases, the difference in evaluation times also increases for the HD-correlated case.

Empirically, we do not find that all computations currently scale as $\mathcal{O}(N_p^3)$.
By fixing the number of frequency bins used in each model and including pulsars in alphanumeric order, we can compare the matrix inversion cost among models and check how the computations scale with number of pulsars.
There are two separate likelihood implementations: a sparse version that uses \texttt{scipy.sparse.csc} compressed sparse column matrices along with a sparse Cholesky decomposition \texttt{scikit-sparse}~\cite{scikitsparse, suitesparseA, suitesparseB} and a version that uses a dense Cholesky decomposition.
Through profiling, we find that the dense computation indeed scales as $\mathcal{O}(N_p^3)$ and is generally slower than the sparse method on a single core.
The sparse method instead scales as $\mathcal{O}(N_p^2)$ using the fast likelihood, \autoref{eq:fast_like}, or as $\mathcal{O}(N_p^{2.3})$ when using the slower version of the likelihood, \autoref{eq:trad_like}.

This $\mathcal{O}(N_p^{2.3})$ empirical scaling of the traditional likelihood displayed in \autoref{fig:corr} can be traced to the sparse Cholesky inversion $\vect{\Sigma}^{-1}$.
In general, the sparse Cholesky inversion does not have a strict scaling, but depends on how the matrix is structured.
Because of this, the scaling of the sparse Cholesky inversion may have some dependence on the number of frequencies and pulsars used.
By profiling the code to see which parts use the largest fraction of the total evaluation time, we find that in the sparse cases, changing from a dense to a sparse matrix takes up the majority of the computation time, and the inversion is performed very quickly by comparison.
This means that our inversion has been sped up to a point where it is no longer the bottleneck of the computation, and the $\mathcal{O}(N_p^2)$ pieces of the computation dominate the overall scaling for \autoref{eq:fast_like} evaluations.
Future work will aim to reduce the computation time required in these $\mathcal{O}(N_p^2)$ operations, perhaps by working with sparse matrices from the start.

Indeed, this increased speed will be important in future analyses.
The addition of pulsar specific noise models to the analysis may require an increase in the number of frequencies $N_\text{freq}$ used in the Fourier basis.
Increasing the number of frequencies means that the bottleneck of the calculation, the matrix inversion, now takes even longer.
With a hundred pulsars and additional frequencies included, a single likelihood evaluation can take seconds to evaluate.
GWB analyses currently require millions of samples, implying that without the faster likelihood, a single analysis would take a significant fraction of a year or more to complete.

\section{Bayesian Methods and Sampling the PTA Parameter Space}
\label{sec:Bayes}
GW data analyses over the past decade have made use of both Bayesian and frequentist statistical techniques.
Bayesian methods rely on Bayes' theorem for parameter estimation and model selection,
\begin{equation}
    p(\vect{\theta} \mid \vect{y}) = \frac{p(\vect{\theta}) p(\vect{y} \mid \vect{\theta})}{p(\vect{y})}\,.
\end{equation}
Given a descriptive model of the relevant data as a likelihood $p(\vect{y}\mid\vect{\theta})$ and a prior distribution $p(\vect{\theta})$ on the parameters, we can sample the (unnormalized) posterior $p(\vect{\theta} \mid \vect{y})$, thus reallocating credibility from the prior across parameter values.
The normalizing factor in the denominator is known as the marginal likelihood or evidence, and it may be used in deciding which of a set of models is preferred.

Numerical methods to estimate $p(\vect{\theta} \mid \vect{y})$ when the dimension of $\vect{\theta}$ is large typically depend on a form of stochastic sampling to perform high-dimensional integrals for both parameter estimation and model selection.
In NANOGrav, we use \texttt{PTMCMCSampler}~\cite{ptmcmcsampler}, a parallel-tempering (PT) enabled Metropolis-Hastings MCMC sampler.
This has been the NANOGrav sampler of choice for many years.
It has also recently been compared to other options and recommended for use in PTA data analysis~\cite{Samajdar:2022qhm}.
To sample the high-dimensional PTA parameter space, we use the default proposal distributions that come with \texttt{PTMCMCSampler}, and some other custom proposals which will be enumerated and described below.
By default, \texttt{PTMCMCSampler} uses single-component adaptive metropolis (SCAM), adaptive metropolis (AM), and differential evolution (DE) proposal distributions, referred to in the code as ``jump proposals.''
Additionally, the sampler supports adding custom distributions.
Details of this sampler were also discussed in~\cite{arzoumanian_nanograv_2014}, but we reiterate them here with any changes that have since been made.

\subsection{Metropolis-Hastings Algorithm}
We use the Metropolis-Hastings algorithm~\cite{metropolis_equation_1953, hastings_monte_1970} to stochastically sample the posterior distribution given by the likelihood and chosen priors.
After enough iterations, the sampler converges to a stable distribution that approximates the posterior well regardless of where in the parameter space we start.
We will show tests of this in \autoref{sec:bayes_tests}.
Under certain circumstances convergence is guaranteed with an infinite number of samples, but in the high-dimensional parameter space that we search over, the amount of time required for convergence can be prohibitive without well-chosen proposal distributions.

First, we start with an initial set of parameters $\theta$ at iteration $t=0$.
Then, we draw a new point $\vect{\theta}_*$ from a proposal distribution for each iteration $J_t\left(\vect{\theta}_*, \vect{\theta}\right)$.
In the Metropolis algorithm~\cite{metropolis_equation_1953}, such proposal distributions are required to be symmetric, but the Metropolis-Hastings (MH) algorithm~\cite{hastings_monte_1970} allows for asymmetric distributions by the inclusion of the Hastings ratio.
The Metropolis-Hastings algorithm leads to the proposed point being accepted with log probability~\cite{gelman_bayesian_2013}
\begin{equation}
    \ln A = \min\left( 0, \ln R + \ln H \right)\,,
\end{equation}
where $R$ is the ratio of probabilities between the proposed point and the old one,
\begin{equation}
    \ln R = \ln p(\vect{\theta}_* \mid \vect{y}) - \ln p(\vect{\theta} \mid \vect{y})\,,
\end{equation}
and $H$ is a ratio which accounts for asymmetric proposal distributions
\begin{equation}
    \ln H = \ln J_t(\vect{\theta} \mid \vect{\theta}_*) - \ln J_t(\vect{\theta}_* \mid \vect{\theta})\,.
\end{equation}
Iterating for each $t$ returns a chain of samples from the posterior distribution.
However, the number of samples required to settle into a stationary distribution that approximates the posterior well depends on several factors including the auto-correlation length, which is related to the number of parameters, and whether the distribution is multi-modal where the chain might get stuck in a single mode.
At each iteration, we draw a point from the proposal distributions which is either accepted or rejected.
If the new point is accepted, it is added to the chain.
Otherwise, if the new point is rejected, the current point is added to the chain instead.
Chains produced in this way contain samples which are not completely independent, and they are correlated with themselves.
In the case of a MH sampler, the auto-correlation length is typically of the same order as the number of parameters.

\subsection{Parallel Tempering}
To improve exploration and mixing of the sampler, we sample multiple chains with different exponents, known as temperatures, and propose swaps between them in a sampling scheme known as PT~\cite{neal_sampling_1996}.
The posterior now contains the likelihood raised to some power,
\begin{equation}
    p\left(\vect{\theta} \mid \vect{y}, \beta\right) = p\left(\vect{\theta}\right) p\left(\vect{y} \mid \vect{\theta}\right)^\beta\,,
\end{equation}
where $\beta = 1 / T$ and $T$ is known as the chain's temperature.
Samples from one chain are propagated to the next via swap proposals between chains of different temperatures.
The higher the temperature is, the more the posterior becomes like the prior, enabling exploration of the parameter space and reducing the auto-correlation length of all chains via swaps, increasing the number of effectively independent samples.
Though the cost of this scheme is more evaluations of the likelihood, we often find that this is more efficient because of the increased number of effectively independent samples returned.
It can also be used to find the evidence as will be discussed in a later section.
We set a temperature for each of the chains using a geometrically spaced ladder,
\begin{equation}
    T_i = \left(\frac{T_{\text{max}}}{T_{\text{min}}}\right)\exp(i)\,,
\end{equation}
where $i=0, 1, 2, \ldots$, which should result in a $\sim25\%$ temperature swap acceptance rate between adjacent chains when sampling a multi-variate Gaussian distribution~\cite{arzoumanian_nanograv_2014}.
In \texttt{PTMCMCSampler}, swaps are proposed between adjacent chains, though in general swaps could be proposed between any two chains in the ladder.
Swaps are accepted between temperatures $T_i$ and $T_j$ with log probability,
\begin{equation}
    \ln A_{i j} = \min \left( 0, \left(\beta_j - \beta_i\right) \ln L_{i j}\right)\,,
\end{equation}
where
\begin{equation}
    \ln L_{i j} = \ln p(\vect{y} \mid \vect{\theta}_{i}) - \ln p(\vect{y} \mid \vect{\theta}_j)\,,
\end{equation}
is the log likelihood ratio.
Other temperature ladders are possible including adaptive temperature spacing based on a constant acceptance rate~\cite{vousden_dynamic_2016}.
This improves the PT scheme when sampling a posterior which is not a multi-variate Gaussian.
While these different temperature spacings certainly have advantages, they are not currently implemented in \texttt{PTMCMCSampler}.

To perform parallel tempering swaps, we use multiple cores to sample each chain simultaneously through the use of \texttt{MPI}~\cite{mpi40} and \texttt{MPI4Py}~\cite{dalcin_mpi4py_2021, dalcin_mpi_2005}.
Previously, the sampler used an asynchronous model for the temperature swaps, so that chains could sample at their own paces and swap as soon as the next one down reaches a specified interval.
These processes are now synchronized using blocking commands, which are necessary for the standard product space sampling method that NANOGrav uses for model selection~\cite{lodewyckx_tutorial_2011} (see Appendix~\ref{sec:syncsampler} for a discussion of the necessity of synchronizing the sampler).

\subsection{Jump Proposals}
The GWB analysis uses several proposal distributions.
These distributions are critical for timely convergence and exploration of the parameter space.
Ideally, jump proposals match the posterior closely to minimize the auto-correlation length of the chain and thus reduce the number of samples that need to be taken.
The combination of all of these proposals has proven to work well for the problem at hand, even with the $\mathcal{O}(100)$ parameters that we work with in the 15-year data set.
The proposal distributions discussed here consist of the default in \texttt{PTMCMCSampler} along with empirical distributions and prior draws.

\subsubsection{Adaptive Metropolis}
Upon initializing the sampler, \texttt{PTMCMCSampler} takes a list of ``parameter groups'' as an argument.
If we believe that multiple parameters will be correlated, we can add them as a group, and the sampler will propose jumps in this subspace.
The full sample space of all parameters together is always a group regardless of new groups.
However, if no groups are given, the entire sample space is considered the only group.
Typically, power law amplitude and spectral indices are grouped together with a sampling group for each individual pulsar.

\texttt{PTMCMCSampler} also requires a sample covariance matrix at initialization.
Periodically, we compute a sample covariance matrix $\vect{C_s}$ from the chain's history for each sample group using an online algorithm (see Appendix~\ref{app:online_alg}) to avoid storing the entirety of the chain.
$\vect{C_s}$ is then decomposed via a singular-value decomposition (SVD),
\begin{equation}
    \vect{C_s} = \vect{U_s}\vect{\Sigma_s} \vect{V_s}^T\,.
\end{equation}
This provides a robust generalization of the eigenvalue decomposition and is used to jump along correlated directions in the parameter space.

The adaptive Metropolis proposal distribution~\cite{haario_adaptive_2001} uses the sample covariance matrix to propose jumps in uncorrelated directions of the parameter space.
First, we move the usual parameters at the $i$th step of the chain $\vect{\theta}_i$ into the parameter combinations using
\begin{equation}
    \vect{\zeta}_i = \vect{U_s}^T \vect{\theta}_i\,.
\end{equation}
Each jump proposed is given by a normal distribution,
\begin{equation}
    \vect{\zeta}_{i + 1} = \vect{\zeta}_{i} + \sqrt{\vect{\Sigma_g}}\,\mathcal{N}(0, c_d)\,,
\end{equation}
where $\vect{\Sigma_g}$ is the sample covariance matrix for the specific group in which the jump is proposed, and
\begin{equation}
    c_d = \frac{2.4 s}{\sqrt{2 n_{\text{dim}}}}\,,
\end{equation}
where $s$ is a scale parameter and $n_{\text{dim}}$ depends on the number of dimensions of the group that the jump is proposed in..
For each default jump, $3\%$ of jumps have their scale multiplied by 10, $7\%$ of jumps have their scale multiplied by 0.2, and the other $90\%$ have unmodified scale.
In all cases, the relative scale of the jump is adjusted based on the temperature of the chain.
The mix of small, medium, and large jumps helps the sampler find the scale of the parameter space being explored.
To project back into a jump in the original parameters, we use
\begin{equation}
    \vect{\theta}_{i + 1} = \vect{U} \vect{\zeta}_{i + 1}\,.
\end{equation}

\subsubsection{Single-Component Adaptive Metropolis}
Similar to the adaptive Metropolis jump, the single-component adaptive Metropolis (SCAM) jump~\cite{haario_componentwise_2005} uses the sample covariance matrix, but only moves along one \textit{uncorrelated} direction in the parameter space.
Once again, we start by projecting onto the uncorrelated combinations of parameters $\vect{\zeta} = \vect{U}^T \vect{\theta}$.
We then propose a jump in a single parameter direction as,
\begin{equation}
    \zeta_{i+1}^j=\zeta_i^j+\sigma_s^j \mathcal{N}\left(0, c_d\right)\,,
\end{equation}
where $\sigma_s^{j}$ is the $j$th diagonal element of the diagonal matrix $\sqrt{\vect{\Sigma_s}}$, and $j$ labels the uncorrelated parameter for which we are proposing a new point.
Finally, we move back to the original set of parameters using
\begin{equation}
    \vect{\theta}_{i + 1} = \vect{U} \vect{\zeta}_{i + 1}\,.
\end{equation}

\subsubsection{Differential Evolution}
The final default proposal distribution in \texttt{PTMCMCSampler} uses a simple genetic algorithm known as differential evolution (DE)~\cite{braak_markov_2006}.
This algorithm takes two samples from the history of the chain, subtracts them, and proposes a jump along that direction.
In the current version of \texttt{PTMCMCMSampler}, the full chain is not stored, and it instead draws from a buffer formed over many unthinned iterations of the sampler.
By keeping this buffer much longer than the auto-correlation length of the sampler, we ensure that the draws come from a stationary distribution.
The DE jump draws two samples $\vect{\theta}_m$ and $\vect{\theta}_n$ and then proposes a jump
\begin{equation}
    \vect{\theta}_{i+1}=\vect{\theta}_i+s_{DE}\left(\vect{\theta}_m - \vect{\theta}_n\right)\,,
\end{equation}
where $s_{DE} = 1$ or
\begin{equation}
    s_{DE} \sim \text{Uniform}\left[0, \frac{2.4}{\sqrt{2 \beta n_\text{dim}}}\right]\,,
\end{equation}
each with $50\%$ probability with $\beta = 1/T$ and $n_\text{dim}$ the number of dimensions in which the jump is proposed.

\subsubsection{Empirical Distributions}
Before the GWB analysis, we run a noise analysis on each pulsar individually.
This run includes white noise (EFAC, EQUAD, ECORR) and a power-law intrinsic red noise (amplitude and spectral index).
From each of these runs, we find a posterior for the intrinsic red noise, and we create white noise dictionaries with which to set the white noise values constant during the full analyses. 
Out of these posteriors, we can make 1D or 2D histograms that we can then sample from during the full GWB analysis to propose as new points.
During the creation of these histograms, all bins' counts are incremented by one to allow exploration of the entire prior.
They are then checked to make sure they cover the prior before starting any sampling that includes them so that they do not bias parameter recovery.
Such histograms are known as empirical distributions (see Appendix A in~\cite{NG11.cw}).
Typically, we use 2D empirical distributions on the intrinsic red noise parameters for each pulsar.
This provides excellent proposals if the empirical distributions are somewhat close to the posteriors on the parameters which we propose jumps in.
Empirical proposal distributions reduce the number of samples required to achieve stationarity, sometimes called the burn-in, significantly.
Importantly, empirical distributions are only a small part of our overall mix of proposal distributions which include many proposals that suggest points across the entire prior space.

\subsection{Parameter Estimation}
Parameter estimation provides crucial information for astrophysical inference.
The full multi-dimensional posterior, can be used to find the marginalized posteriors for each of the parameters.
Convergence and exploration of the parameter space is critical to finding the maxima of the likelihood function and to sampling them effectively.
GWB analyses typically involve millions of likelihood evaluations due to the large auto-correlation lengths of the chains.
It can take many evaluations to get a reasonable number of effective samples.
As a test of the procedure, we have verified the different runs on a single simulation return the same result regardless of where we start in the parameter space.

\subsubsection{Gelman-Rubin $\hat{R}$ Diagnostic}
We use the Gelman-Rubin $\hat{R}$ diagnostic~\cite{gelman_inference_1992} to check the stationarity of the chain.
This does not necessarily mean that the chains have converged, but it tells us that the samples are coming from one part of the parameter space.
The diagnostic splits the MCMC chain into multiple segments and checks the within-chain and between-chain statistics to confirm that the chain is in a stationary state.
A threshold is required to tell whether the chain passes or fails the diagnostic.
As suggested by~\cite{vehtari_rank-normalization_2021}, we use 1.01 as the $\hat{R}$ threshold.
In all tests performed in this work, we use the $\hat{R}$ diagnostic to make sure the chains are stationary before performing tests.
This diagnostic is implemented in \texttt{la\_forge} for ease of use in PTA data analysis.

\subsection{Model Selection: Bayes Factor Calculations}
To compare models, we need to compute the marginal likelihood or evidences for each model.
Division of these evidences return Bayes factors.
The standard method of calculation used in NANOGrav is a form of product space sampling~\cite{carlin_bayesian_1995,lodewyckx_tutorial_2011}.
We refer to this method as the \texttt{HyperModel} (henceforth hypermodel) framework, and it is implemented in \texttt{enterprise\_extensions}.
Other methods for finding the model evidence include reweighting the posterior~\cite{hourihane_accurate_2023-1}, thermodynamic integration~\cite{ogata_monte_1989,gelman_simulating_1998}, and nested sampling~\cite{Buchner2021}.

\subsubsection{HyperModel Framework}
The hypermodel concatenates different models into a single model by combining their joint parameters into a set of parameters that contains only the unique parameters between the models.
During sampling, a continuous ``switch'' parameter called \texttt{nmodel} changes between the models turning on only the parameters that belong to the ``on'' model.
By sampling the models and this switch between them, we can compute an odds ratio comparing how many times each of the models were sampled.
This corresponds directly to a Bayes factor between the two models.
The uncertainties are then computed using a standard bootstrap in which we resample the thinned \texttt{nmodel} marginalized posterior with replacement and recompute the odds ratio.
The odds ratio is averaged and a standard deviation calculated over a number of realizations to give the final Bayes factor \footnote{Note that the odds ratio is only equal to the Bayes factor if the prior odds are the same for each model in consideration.} with uncertainties.

In current data, we often find the situation where Bayes factors for one model or another are significantly disfavored.
Adding a log weight to the log likelihood can remedy such a situation.
To accomplish this, we add a constant value to the log likelihood, scaling the likelihood by the exponential of the log weight.
We can find a weight estimate by subtracting the maximum posterior values between the two models being compared.
This results in a more even mixing between the two models, and the weight can be undone in post-processing by multiplying the Bayes factor by the exponential of the log weight.

\subsubsection{Reweighting}
Reweighting is a simple technique that utilizes existing samples from a probability distribution, the approximate, to obtain an estimate of some other probability distribution, the target, which shares support with the approximate.
Each existing sample is ``weighted" by the ratio of the target and approximate probability densities.
These weighted samples are an estimate of the target distribution whereas the weights can be used to estimate Bayes Factors and uncertainties.
Since the samples of the distribution have already been produced, each weight can be calculated in parallel, increasing the speed at which the target space can be evaluated.
Reweighting results in a reduction in the number of effectively independent samples based on how disjoint the posteriors are between the approximate and target, but it often is still much faster than directly sampling the target posterior directly.
This technique is particularly effective in cases where the two distributions have similar support on similar regions in the parameter space but one distribution is significantly faster to evaluate.
In the context of PTAs, reweighting has been used to generate HD posteriors from the faster-to-evaluate CURN posteriors~\cite{hourihane_accurate_2023-1}.

\subsubsection{Nested Sampling}
Nested sampling~\cite{skilling_nested_2006}, a staple for GW analyses in current ground-based detectors, returns a Bayes factor and posterior samples.
It computes the evidence integral by turning the multi-dimensional integral into a one-dimensional integral.
$N$ ``live'' points are sampled from the priors and the space outside of the lowest likelihood point is removed, thereby reducing the volume considered by approximately $1/N$.
In this way, nested sampling climbs the likelihood distribution in a global way, and eventually reaches a stopping criterion set by the error on the log evidence.
Overall it has a reputation for being easy to use, good at finding multi-modality, and has stopping conditions that do not require much input from the user~\cite{buchner_nested_2023}.
In PTA data, we find it difficult to use nested sampling on the full GWB analysis due to the high number of parameters.
However, we use nested sampling with a reduced set of the data below to check our Bayes factors with $30$ parameters.
In this case, using \texttt{Ultranest}~\cite{Buchner2021} returns the evidence for each model, but with much larger uncertainty over the same computation time span as the hypermodel.

\subsubsection{Thermodynamic Integration}
If we use enough chains over a broad enough set of temperatures, parallel tempering also can be used to compute the evidence.
By taking the average of the log likelihood on each sampled temperature, we can integrate over this to yield the model evidence.
The log evidence is given by~\cite{ogata_monte_1989,gelman_simulating_1998},
\begin{equation}
    \label{eq:ti_ev}
    \ln p(\vect{\delta t} \mid M)=\int_{-\infty}^0 \beta \mathrm{E}_\beta[\ln p(\vect{\delta t} \mid \vect{\eta}, M)] d \ln \beta\,,
\end{equation}
where $\beta = 1 / T$, $M$ is the model of interest, and we are integrating with respect to $\ln \beta$.
We use two separate methods to evaluate uncertainties on evidence estimates.
In one, we use a cubic spline which is fit using a trans-dimensional algorithm known as reversible-jump MCMC~\cite{cornish_bayeswave_2015-1}, and in the other we use a bootstrap on an interpolation between points in the integrand of \autoref{eq:ti_ev}.
There are two types of uncertainty here: the discretization error which is determined by the number of temperatures that we choose, and the sampling error which is determined by the number of independent samples.

\section{Frequentist Methods and the Optimal Statistic}
\label{sec:frequentist}
Fully Bayesian methods, especially when used for model selection, can be computationally expensive.
In this section, we consider a detection statistic and an estimator for the GWB that are built from directly cross-correlating arrival times between pulsars.
We first consider the statistic in the case where the amplitude of the GWB is small compared to the noise, which is what has traditionally been assumed. This statistic is still important for null hypothesis testing, although it can be improved upon when used as an estimator for the strength of the background.
We then move on to consider the situation where the background cannot be neglected, and present an estimator that properly accounts for the background itself. We also discuss how to construct a ``binned'' estimator across the sky to yield a HD reconstruction that takes into account the size of the background. We finish by presenting a version of the optimal statistic (OS) that simultaneously fits for multiple correlation patterns.  We also highlight the effect of the choice of noise parameters used in constructing the OS.

\subsection{Traditional optimal statistic}

We begin by considering the noise-weighted match between the correlation of data in pulsar $a$ with data in pulsar $b$:
\begin{align}
    \label{eq:paired_correlation}\rho_{ab} =& \frac{ \vect{\delta t}_a^T \vect{\mathbf{P}}_a^{-1} \tilde{\vect{\Phi}}_{a b} \mathbf{P}_b^{-1} \vect{\delta t}_b}{\operatorname{tr}\left(\mathbf{P}_a^{-1} \tilde{\vect{\Phi}}_{a b} \mathbf{P}_b^{-1} \tilde{\vect{\Phi}}_{b a}\right)}
    \equiv \vect{\delta t}_a^T \vect{Q}_{a b}\,\vect{\delta t}_b\,,\\
    \vect{Q}_{ab} =& \frac{\mathbf{P}_a^{-1} \tilde{\vect{\Phi}}_{a b} \mathbf{P}_b^{-1}}{\operatorname{tr}\left(\mathbf{P}_a^{-1} \tilde{\vect{\Phi}}_{a b} \mathbf{P}_b^{-1} \tilde{\vect{\Phi}}_{b a}\right)},
\end{align}
where for two different pulsars, $a$ and $b$, we have
\begin{align}
    \label{eq:c_matrix_os}\mathbf{P}_a &= 
    \left< \vect{\delta t}_a \vect{\delta t}_a^T \right> = \vect{D}_a + \vect{F}_a \vect{\phi}_{aa} \vect{F}_a^T\,,\\
    \label{eq:s_matrix_os}\vect{\tilde \Phi}_{ab} &=\frac{\vect{F}_a\vect{\phi}_{ab}\ \vect{F}_b^T}{\Gamma_{ab}A^2_{gw}}\,.
\end{align}
The normalization of $\vect{\tilde \Phi}_{ab}$ is chosen such that $\langle \rho_{ab}\rangle = \Gamma_{ab}A^2_{\textrm{gw}}.$ In the small signal regime, the variance of this correlation is $\sigma_{ab}^2 = \left(\operatorname{tr}\mathbf{P}_a^{-1} \tilde{\vect{\Phi}}_{a b} \mathbf{P}_b^{-1} \tilde{\vect{\Phi}}_{b a}\right)^{-1}.$ If we assume that the cross-correlation for one pair of pulsars is not correlated with the cross-correlation of another pair of pulsars, i.e., $\langle \rho_{ab}\rho_{cd}\rangle \propto \delta_{ac}\delta_{bd}$, then we can perform a variance-weighted, HD-matched sum over these correlations to estimate the amplitude of the GWB from all pairs,
\begin{align}
\label{eq:a_gw_estimator_no_correlations}\hat A_{\textrm{gw}}^2 &= \frac{\sum_{a} \sum_{b>a} \rho_{ab} \Gamma_{ab} \sigma_{ab}^{-2}}{\sum_{a} \sum_{b>a}\Gamma_{ab}^2\sigma_{ab}^{-2}}\,,\\
\label{eq:sigma_gw_estimator_no_correlations}\sigma_{\textrm{gw}}^2 &= \left(\sum_a\sum_{b>a}\sigma_{ab}^{-2}\Gamma_{ab}^2\right)^{-1}.
\end{align}

This statistic is often used for null hypothesis testing for GWB detection. Under the (null) assumption that $A^2_{\textrm{gw}}=0$, the variance of this estimator can be used to construct a signal-to-noise ratio (S/N),
\begin{align}
\label{eq:snr_os_no_correlations}\textrm{S/N} &= \frac{\sum_{a} \sum_{b>a} \rho_{ab} \Gamma_{ab} \sigma_{ab}^{-2}}{\left(\sum_{a} \sum_{b>a}\Gamma_{ab}^2\sigma_{ab}^{-2}\right)^{1/2}}\,,
\end{align}
which we calculate on the data and then compare to its expected distribution under the null hypothesis. The distribution for this statistic is a generalized $\chi^2$ distribution~\cite{Hazboun:2023tiq}, but it is often estimated empirically using methods that destroy correlations, but preserve potential mismodelling. Two such methods are sky scrambles~\cite{Cornish:2015ikx} and phase shifts~\cite{Taylor:2016gpq}.

Construction of the matrices in \autoref{eq:c_matrix_os} and \autoref{eq:s_matrix_os} requires a choice of hyperparameters $\vect{\eta}$.  Specifically, the red noise parameters for each pulsar are used to construct $\vect{D}_a$, and the CURN amplitude and spectral index are used to construct $\vect{\phi}_{ab}$. The natural choice for these parameters are those taken from the Bayesian analysis. However, choosing the maximum likelihood parameters for $\vect{\eta}$ from a fully Bayesian run, or jointly maximizing the individual 1-D posteriors for each parameter, leads to a bias in the recovered value of $A_{\textrm{gw}}^2$~\cite{vigeland_noise-marginalized_2018}. Part of this bias is due to making a single choice of noise parameters.
By averaging the statistic calculated over draws of $\vect{\eta}$ from a posterior chain, resulting in what is referred to as the ``noise marginalized optimal statistic'' (NMOS) this bias can be partially alleviated. Additionally, a single choice of hyperparameters could result in a larger value of the S/N than is representative of the dataset. In general, therefore, the S/N is averaged over many draws from $p(\vect\eta|\vect{\delta t})$, and this average is used as a detection statistic. Other approaches have also been proposed, e.g., averaging the $p$-value associated with the S/N for each draw, instead of averaging the S/N~\cite{ppc1}.

The OS defined in \autoref{eq:a_gw_estimator_no_correlations} and \autoref{eq:sigma_gw_estimator_no_correlations}.
It is implemented in the \texttt{compute\_os} method of the \texttt{OptimalStatistic} class, which is found in the \texttt{frequentist.optimal\_statistic} module of \texttt{enterprise\_extensions}. The NMOS (described in the previous paragraph) is implemented as the \texttt{compute\_noise\_marginalized\_os} of the same class, and takes a MCMC chain and a list of parameter names as input.

\subsection{Optimal statistic with a non-negligible GWB}
In the case where $A_{\textrm{gw}}^2$ is comparable to the red noise level in some pulsars, the assumption that $\langle \rho_{ab}\rho_{cd}\rangle \propto \delta_{ac}\delta_{bd}$ breaks down. We must account for the covariance between correlations when constructing both $\hat A^2_{\textrm{gw}}$ and especially its variance, which will be dominated by the background itself. When accounting for the covariance between correlations due to the GWB we find
\begin{align}
\vect{\Sigma_{ab,cd}} =& \langle\rho_{ab}\rho_{cd}\rangle - \langle\rho_{ab}\rangle\langle\rho_{cd}\rangle \label{eq:rho_covariances}\\
=& \nonumber
\operatorname{tr}\left(
\vect{Q}_{ba}\mathbf{P}_{ac}\vect{Q}_{cd}\mathbf{P}_{db}\right) \\&+ \operatorname{tr}\left(\vect{Q}_{ba}\mathbf{P}_{ad}\vect{Q}_{dc}\mathbf{P}_{cb}\right)\,,\end{align}
where
\begin{align}
    \mathbf{P}_{ab} &= 
    \left< \vect{\delta t}_a \vect{\delta t}_b^T \right> = \delta_{ab}\vect{D}_a + \vect{F}_a \vect{\phi}_{ab} \vect{F}_b^T\,,
\end{align}
We can then construct a least-squares estimator for the background using this covariance matrix,
\begin{align}
\hat A_{\textrm{gw}}^2 &= \frac{ \bm \Gamma^T\bm \Sigma^{-1} \bm\rho}{\bm \Gamma^T \bm \Sigma^{-1} \bm \Gamma}\,,\label{eq:estimator_with_correlations}\\
\label{eq:sigma_gw_with_correlations}\sigma_{A_{\textrm{gw}}^2}^2 &= \left(\bm \Gamma^T \bm \Sigma^{-1} \bm \Gamma\right)^{-1}\,,
\end{align}
where $\bm \Sigma$ is given by \autoref{eq:rho_covariances}, $\bm\rho$ is a vector of paired correlations and $\bm \Gamma$ is a vector of the HD correlation coefficients corresponding to each pair. It is important to note here that construction of $\mathbf{P}_{ab}$ and therefore $\vect\Sigma$ requires a choice of $\vect\eta$. That is, some choice of $A^2_{\textrm{gw}}$ is needed to actually construct our estimator. One can take an iterative approach, where we begin with a choice of $\vect\eta$ [e.g., the maximum \textit{a posteriori} draw from $p(\vect\eta|\vect{\delta t})$], evaluate \autoref{eq:estimator_with_correlations}, and then use the resulting $\hat A^2_\textrm{gw}$ to construct $\vect\Sigma$, iterating until convergence. In practice, we have found this converges rapidly, and is consistent with results that use a single iteration with an initial choice of $A^2_\textrm{gw}$ estimated from the posterior $p(\vect\eta | \vect{\delta t})$.

It is also common to estimate the strength of the observed background using subsets of paired correlations that have similar separations on the sky. The individual bins should then trace the HD curve. We collect pairs that fall into a single bin, initially choosing the number of bins we would like, and then assigning pulsar pairs to bins such that there are roughly the same number of pairs in each bin.  We label the average angular separation between pulsars in the $i^\textrm{th}$ bin as $\xi_i$, and construct an estimator for the correlated power in each bin, $\rho_{\textrm{opt,}i}$, whose expectation is given by $\langle\rho_{\textrm{opt,}i}\rangle = \Gamma_{\xi_{i}} A_{\textrm{gw}}^2$. In this case $\Gamma_{\xi_i}$ is the HD correlation coefficient evaluated at the average angular separation for pulsar pairs in the $i^\textrm{th}$ bin. Other choices could also be made, and would slightly change the results~\cite{Allen:2022ksj}.
 Motivated by our choice of binning, we can imagine $\vect{\Sigma}$ taking a \textit{block form} where the $i,i$ block corresponds to correlations between pairs in bin $\xi_{i}$, the $i,j$ block corresponds to correlations between pairs in $\xi_{i}$ and pairs in $\xi_{j}$. The resulting estimator is given by~\cite{Allen:2022ksj}
\begin{align}
    \label{eq:binned_estimator_with_covariances}
    \rho_{\textrm{opt,}i} = \Gamma_{\xi_i}\frac{\vect{\Gamma}_{i}^T\vect{\Sigma}_{ii}^{-1}\vect{\rho}_{i}}{\vect{\Gamma}_{i}^T\vect{\Sigma}_{ii}^{-1}\vect \Gamma_{i}},
\end{align}
where $\vect\Gamma_i$ is a vector of the overlap reduction function for all pairs in bin $i$, $\vect\rho_i$ is the vector of paired correlations in bin $i$. The non-zero GWB induces correlations between these bins as well, and this covariance matrix is given by~\cite{Allen:2022ksj}
\begin{align}
    \label{eq:binned_estimator_covariance_matrix}
    B_{ij} &= \langle \rho_{\textrm{opt,}i} \rho_{\textrm{opt,}j}\rangle - \langle \rho_{\textrm{opt,}i}\rangle \langle \rho_{\textrm{opt,}j}\rangle, \\
    &= \Gamma_{\xi_i}\Gamma_{\xi_j}\frac{\vect \Gamma_{i}^T \vect{\Sigma}_{ii}^{-1}\vect{\Sigma}_{ij}\vect{\Sigma}_{jj}^{-1}\vect \Gamma_{j}}{\left(\vect \Gamma_{i}^T \vect{\Sigma}_{ii}^{-1}\vect \Gamma_{i}\right)\left(\vect \Gamma_{j}^T \vect{\Sigma}_{jj}^{-1}\vect \Gamma_{j}\right)}\,.
\end{align}

\subsection{Optimal statistic for multiple correlation patterns}
Reference~\cite{sardesai_generalized_2023} arrived at the multiple component optimal statistic (MCOS) through a $\chi^2$ approach, with
\begin{align}
    \chi^2 = \sum_{ab}\left(\frac{\rho_{ab} - \sum_\alpha A^2_\alpha \Gamma_{ab}^\alpha}{\sigma_{ab}}\right)^2,
\end{align}
$\rho_{ab}$ given by \autoref{eq:paired_correlation}, and $\sigma_{ab}$ its associated uncertainty. The label $\alpha$ now indexes different spatial correlation patterns. For example, one can jointly minimize $\chi^2$ with respect to $A^2$ for both a HD-correlated process and a monopole process, and calculate the associated covariance matrix between those estimators.  

By generalizing the optimal statistic to include more than one ORF simultaneously, one arrives at the \textit{multiple-correlation} optimal statistic (MCOS),
\begin{equation}
\label{eq:mcos_estimator}
    \hat{A}_\alpha^2 = \sum_\beta B_{\alpha \beta} C^\beta\,,
\end{equation}
where $\alpha$ and $\beta$ are individual ORFs, and
\begin{align}
    B^{\alpha \beta} &\equiv \sum_{a} \sum_{b > a} \frac{\Gamma_{a b}^\alpha \Gamma_{a b}^\beta}{\sigma_{a b}^2} \,,\\
    C^\beta &\equiv \sum_{a} \sum_{b > a} \frac{\rho_{a b} \Gamma_{a b}^\beta}{\sigma_{a b}^2}\,.
\end{align}
When denoting matrices in this notation, upper and lower indices indicate matrix inverses with respect to one another. 
The variance on the individual estimators for each spatially-correlated process is given by
\begin{equation}
\label{eq:mcos_estimator_sigma}
    \sigma_{\hat{A}_\alpha^2}^2=B_{\alpha \alpha}\,.
\end{equation}
The noise-marginalized version of the MCOS follows a similar structure as the noise-marginalized version of the original optimal statistic.
By drawing parameters from the MCMC chains, we average over the noise and return a distribution of values.

The MCOS is implemented in the \texttt{compute\_multiple\_corr\_os} method of the \texttt{OptimalStatistic} class. The noise marginalized version of the MCOS is implemented as the \texttt{compute\_noise\_marginalized\_multiple\_corr\_os} method.

\section{Tests of Bayesian Methods: PTMCMCSampler and Enterprise}
\label{sec:bayes_tests}
Here, we perform tests robustness of our Bayesian methods.
The tests performed here use the faster likelihood with the two-step marginalization procedure.
We begin these tests by checking prior recovery.
This tests that our proposal distributions satisfy detailed balance, a condition required for the chain samples to reflect the posterior.
Next, we create simulations based on the 15-year NANOGrav data and check for unbiased posteriors.
Simulations are produced directly from \texttt{Enterprise} models using the TOAs from the 15-year data.
Therefore, good posterior recovery implies that the models we use are self-consistent and that the recovered posteriors are in the correct place with the right width.
Finally, we use a reduced version of the data to check that Bayes factors agree among different methods of computation.
If the different methods agree, we conclude that our calculations are working properly.

\subsection{Prior Recovery Tests}
To test the proposal distributions incorporated in \texttt{PTMCMCSampler} and \texttt{enterprise\_extensions}, we sample a posterior that is equal to the priors by setting the likelihood equal to the prior and the prior to a constant.
PT only tempers the likelihood in \texttt{PTMCMCSampler}, so it is necessary to sample the prior as the likelihood to also test this part of the sampler.
If the proposal distributions satisfy detailed balance, then the recovered posterior equals the input priors within sampling uncertainties.
\texttt{PTMCMCSampler} contains three default proposal distributions known inside the code as ``jump proposals.''
Additional jump proposals come from \texttt{enterprise\_extensions}, and which proposals get used depends on the type of search being performed.
The isotropic GWB analyses includes AM, SCAM, and DE proposals.
Additionally, this search includes prior draws and two-dimensional empirical distributions on the intrinsic red noise amplitudes and spectral indices for each pulsar.

To assess prior recovery, we use a quantile-quantile (Q-Q) plot.
Q-Q plots compare the quantiles of the distribution of our recovered prior with samples drawn from a simulated distribution of the input.
We subtract the mean of the simulated distribution from every point in our plot so that the mean falls along zero on the vertical axis.
The expected result is a set of lines, one for each parameter, falling within the given uncertainties with few venturing outside the $3\sigma$ bounds.
Bias could appear as a line remaining significantly above or below the mean for the entire interval indicating that more samples exist in one quantile of the distribution.

\begin{figure*}
    \centering
    \includegraphics[width=\textwidth]{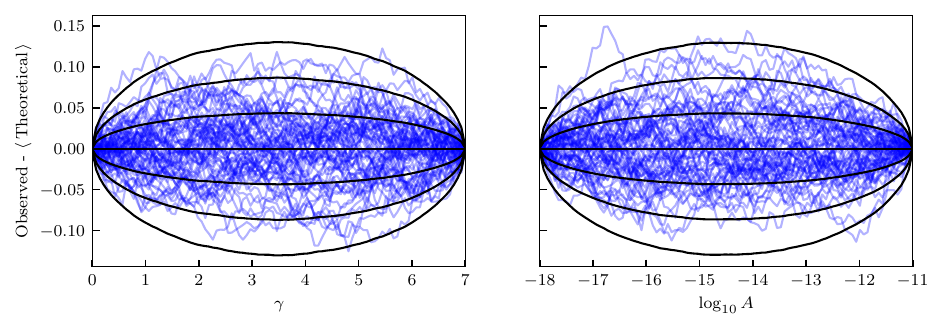}
    \caption{
    Quantile-quantile plot showing the recovery of the prior.
    To produce this plot, we sample the prior in place of the likelihood and set the prior to a constant.
    This is not the same as setting the likelihood to a constant, because we use a parallel tempering method that only tempers the likelihood.
    Priors on the $\gamma$ parameters are Uniform$[0, 7]$, and the priors on the $\log_{10}A$ parameters are Uniform$[-18, -11]$.
    The chains produced should be equal to the input prior distribution within the sampling uncertainties after thinning.
    On the horizontal axis, we plot the parameter value associated with the quantiles of the simulated uniform distribution.
    On the vertical axis, we plot the parameters associated with the quantiles of the distribution output from the sampling process minus the mean of simulated parameters so that the mean lies along the zero of the vertical axis.
    The curved, solid lines show $1\sigma$, $2\sigma$, and $3\sigma$ uncertainties.
    The uncertainty lines are created by taking the average and standard deviation over 10,000 realizations of a uniform distribution with the same number of samples as the observed distribution.
    This plot shows that the priors were recovered correctly when using parallel tempering, AM, SCAM, DE, empirical, and prior proposal distributions.
    }
    \label{fig:qqplot}
\end{figure*}

We use the same prior for each spectral index parameter $\gamma$ and for each log amplitude parameter $\log_{10}A$,
\begin{align}
    \gamma &\sim \text{Uniform}[0, 7]\,,\\
    \log_{10} A &\sim \text{Uniform}[-18, -11]\,.
\end{align}
These priors are chosen specifically for this test and may be slightly different in ``production grade'' analyses between the intrinsic red noise amplitudes and the GWB amplitude, because we can rule out very large GWB amplitudes since they have not been observed in previous data sets.
Spectral index priors are typically the same for all parameters.
The NANOGrav 15-year GWB analysis uses 67 pulsars.
Assuming a power-law spectrum, each pulsar adds a spectral index and an intrinsic red noise amplitude parameter.
Along with the common process spectral index and amplitude parameters, each of the plots in \autoref{fig:qqplot} contains 68 lines.
The majority of the lines remain inside three sigma uncertainty bounds and only occasionally do lines venture outside and then back toward zero.

\subsection{Simulation Recovery Tests}
Next, we create simulations to check that \texttt{PTMCMCSampler} returns unbiased posteriors.
To quantify whether our posteriors are unbiased, we use a P-P plot\footnote{We could use Q-Q plots, which are quite general, for this too. However, P-P plots are somewhat standard among the literature for testing simulation recovery.}.
We make these plots by creating 100 realizations of data, sampling the posteriors, and finding the $p$-value at which the simulated value falls.
These $p$-values should follow a uniform distribution.
By taking the CDF of these values and plotting them against the $p$-values, we expect each parameter (shown as a line on the plot) to follow a diagonal line within some confidence interval.
Since we model the red noise as a power law and have 67 pulsars, there are 136 parameters searched over in each simulation: an amplitude and spectral index for each pulsar's intrinsic red noise and an amplitude and spectral index for the common red process which is shared among all pulsars.
For this P-P plot, we use the CURN model, because the computational expense is small compared to the HD correlated model, and the posteriors between the two models are similar.
Furthermore, \citet{hourihane_accurate_2023-1} found that reweighting the CURN plots using the HD model's likelihood also resulted in diagonal P-P plots.
Given the similarity between the two models' posteriors, \texttt{PTMCMCSampler} should have no extra problems with exploring the HD parameter space given its effective exploration of the CURN parameter space.

\begin{figure}
    \centering
    \includegraphics{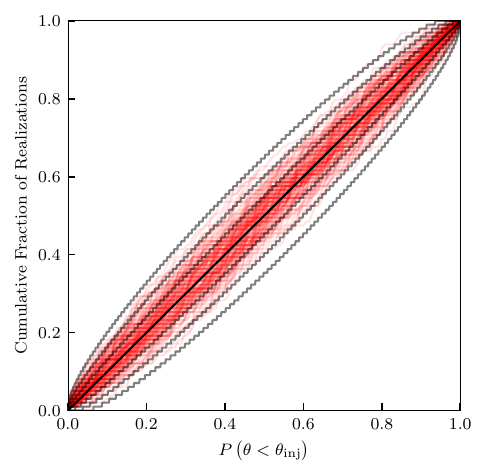}
    \caption{A P-P plot shows the recovery of simulated parameters for 100 realizations simulated using the 15-year NANOGrav data with simulated values pulled from the priors in \autoref{tab:sim_prior}. The model being simulated includes power law intrinsic red noise parameters and a power law CURN. Each of the 136 parameters are presented as an individual line on the P-P plot. A diagonal black line indicates perfect recovery, and the 68.27\%, 95.45\%, 99.73\% confidence intervals, found using an inverse CDF of a binomial distribution~\cite{ashton_bilby-mcmc_2021}, appear as curved black lines with 68.27\% being the closest to the diagonal and 99.73\% being farthest away. The plot indicates no significant bias in recovery of the posterior by \texttt{PTMCMCSampler} when the simulation and recovery is performed via \texttt{Enterprise}. }
    \label{fig:pp_plot}
\end{figure}

We make simulations differently here than in other papers which have used either \texttt{libstempo}~\cite{vallisneri_libstempo_2020} and \texttt{TEMPO2} or \texttt{PINT}.
The question of whether our models match the simulations of realistic PTA data is outside the scope of this study.
Instead, we use the models in \texttt{Enterprise} to simulate pulsar residuals with timing model uncertainties, specified white noise, intrinsic red noise, and either a CURN or a HD correlated GWB.
The priors set on these values are as shown in \autoref{tab:sim_prior}.
Crucially, the distributions which we draw values from and the priors we search over must be the same.
If they are not, the P-P test will fail.
We have reduced the prior space from the full production analyses to limit ourselves to a detectable part of the parameter space.
While this is not required to make good P-P plots, our purpose is to test whether we get unbiased results in the event of a set of parameters which have strong signals visible.
\begin{table}[t]
    \centering
    \begin{tabular}{c|c|c}
        Parameter     & Prior & Values \\
        \hline
        IRN Amplitude & Uniform & $[-15, -12]$ \\
        IRN Spectral Index & Uniform & $[2, 6]$ \\
        CURN Amplitude & Uniform & $[-16, -14]$ \\
        CURN Spectral Index & Uniform & $[2, 6]$ \\
    \end{tabular}
    \caption{Priors sampled for the simulated values of simulation realizations. IRN indicates an intrinsic red noise parameter and CURN indicates a common uncorrelated parameter which is common to all pulsars. These priors are also used in the search of the simulations to recover posteriors for use in P-P tests.}
    \label{tab:sim_prior}
\end{table}

The P-P plot produced in \autoref{fig:pp_plot} indicates little bias in the recovery of the simulated values.
Lines that remain above and below the diagonal could indicate that the recovery is biased so that we find the simulated values either too low or too high consistently.
A signature ``S'' shape in which the line goes above (below) and then below (above) the diagonal indicates that the width of the posterior distribution is over or under estimated.
After checking each of these 136 parameters individually, we find no evidence of either of these issues.

\subsection{Bayes Factor Recovery Tests}
In the final segment of the Bayesian tests, we check how well our Bayes factors are recovered.
Here, we use a few different techniques of computing the Bayes factors and compare between them.
Due to computational limitations, we reduce our data set to a set of 14 pulsars that have been timed for greater than 15 years.
This allows us to use nested sampling which does not converge quickly for high dimensional spaces such as with the full 67 pulsar parameter space.
Simulations were made with varying noise realizations and GWB amplitudes across a broad range of log Bayes factors,
\begin{equation}
    \log_{10}\text{BF} = \log_{10}Z_{\text{HD}} - \log_{10}Z_{\text{CURN}}\,,
\end{equation}
from $-2.5$ to $17$, where $Z$ is the evidence.
Bayes factors are computed for the HD correlated model against the CURN model.
We check the hypermodel using \texttt{PTMCMCSampler} against nested sampling with \texttt{UltraNest}, and the hypermodel results against Bayes factors returned with reweighting~\cite{hourihane_accurate_2023-1}.
By running these methods on the same 100 realizations with each method, we show that they return consistent answers, although at different levels of uncertainty.

In the comparison between reweighting and the hypermodel, we take the resulting chain from the hypermodel and reweight any of the uncorrelated model samples to the HD correlated model.
This gives us a set of weights which can be used to compute a Bayes factor and uncertainties.
We find that the two methods give results on the log ratio that are consistent with zero in every realization within $3\sigma$.
In every case, reweighting gives a larger uncertainty than the hypermodel and dominates in the subtraction of Bayes factors in which errors are propagated in quadrature.

Nested sampling requires a stopping condition in terms of the uncertainty on the log evidence.
Unfortunately, we find that even with the reduced parameter space, setting the stopping condition to $d\log Z < 0.5$ led to week-long run times.
Once again, the uncertainties on the hypermodel are overwhelmed by the uncertainties on nested sampling without requiring more computational resources for nested sampling, 
and all realizations are consistent with zero within $3\sigma$.

\begin{figure*}
    \centering
    \includegraphics{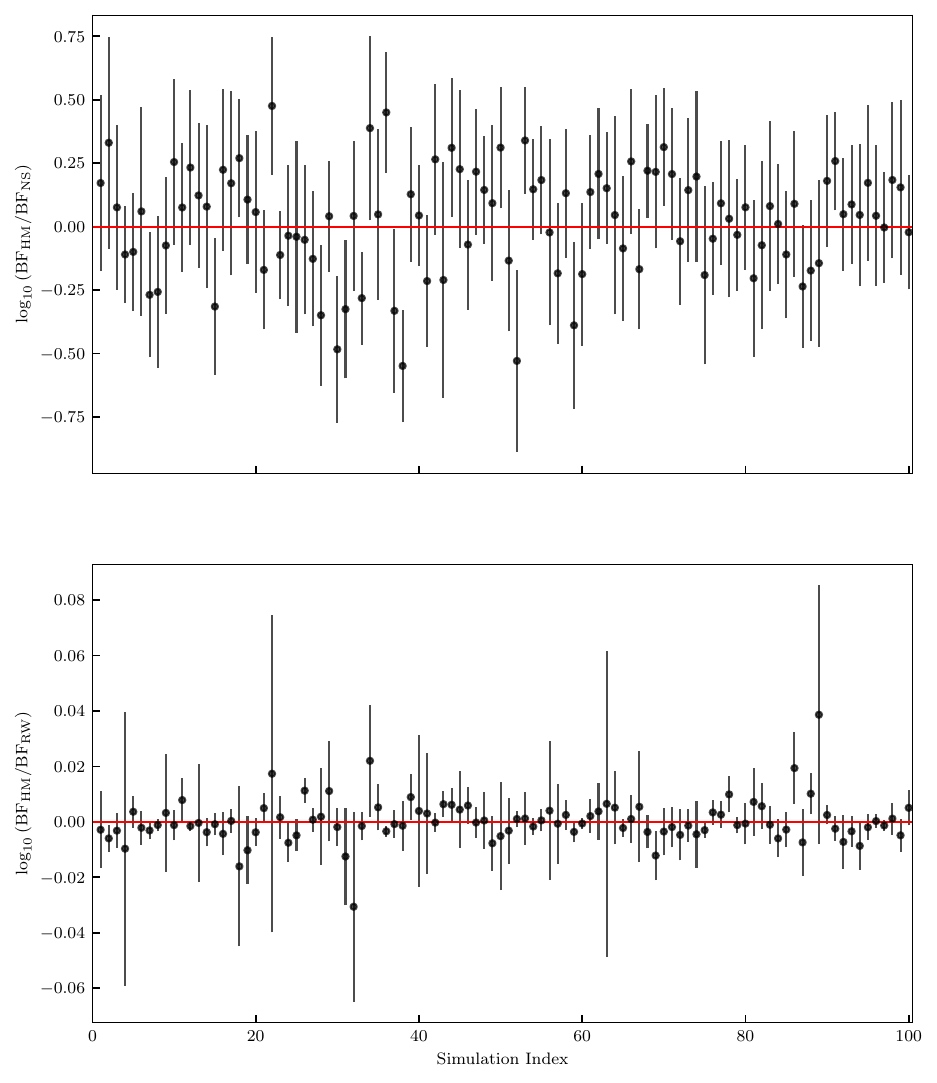}
    \caption{Logarithmic ratio of the Bayes factor computed for each of 100 simulations using the \texttt{HyperModel} framework and nested sampling in the top panel and the \texttt{HyperModel} framework and reweighting in the bottom panel. Each point indicates a mean value of the ratio and the $1\sigma$ uncertainties are given as a vertical bar on each point. The red line indicates zero on the vertical axis, where we expect these values to fall if the Bayes factors returned from each method are consistent. Uncertainties are dominated by the nested sampling and reweighting methods. The values are consistent with zero within $3\sigma$ across all simulations.}
    \label{fig:delta_bfs_hmrw}
\end{figure*}

\begin{figure*}
    \centering
    \includegraphics{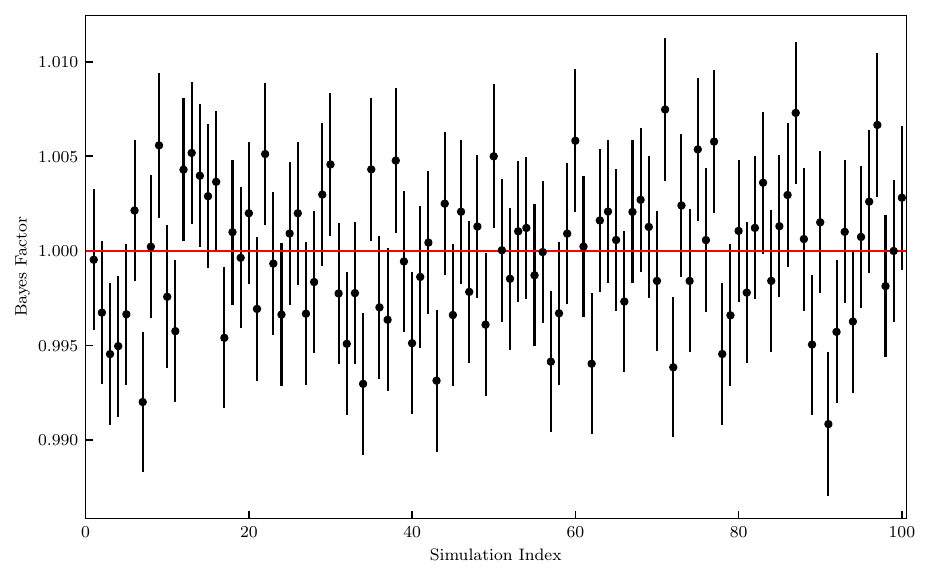}
    \caption{Bayes factors computed for a CURN model against itself for the same 100 simulations as used in the other tests of Bayes factors. The red line in this figure indicates the true Bayes factor between the two models. All realizations recover the true value within $3 \sigma$. This represents a quick check to see if the Bayes factor calculation using a \texttt{HyperModel} returns correct answers in the ideal scenario of posteriors that are exactly the same between the two models being sampled.}
    \label{fig:autobfs}
\end{figure*}

As a final test of the Bayes factors, we run the hypermodel on a CURN model against itself on the same 100 simulations that were used above.
In this case, we know that the Bayes factor must equal one, because a model should not be preferred over itself.
On top of checking whether we get an answer consistent with the known value, this test shows whether the uncertainties are being estimated properly.
As shown in \autoref{fig:autobfs}, the Bayes factor of one is recovered in every realization within $3 \sigma$.
This method represents an easy check that can be performed for any situation to make sure that the hypermodel calculation is working.
However, this test is not sufficient to claim that the method will work for all scenarios.
The case of a model against itself does not take into account the situation where the posteriors between two models are very different.
Therefore, the test of consistency against other samplers remains necessary.

\section{Tests of frequentist methods: enterprise\_extensions and the optimal statistic}
\label{sec:frequentist_tests}

In this section we present a series of tests on the optimal statistic presented in \autoref{sec:frequentist}.
We begin by using simulated data sets to compare the ``traditional'' optimal statistic and the one that accounts for covariance between pulsar pair correlations.
We use those same data sets to evaluate how the revised binned estimator performs as well.
This is followed by a discussion of the distinction between using these statistics as estimators for the amplitude of the GWB vs. using them as detection statistics in a classical null hypothesis testing scenario.
We finish by summarizing recent work in the literature on the multiple component optimal statistic and constructing empirical and analytic distributions for the optimal statistic. 

\subsection{The optimal statistic as an estimator}
\label{subsec:optstat_est}
In the case where $A_{\textrm{gw}}^2$ is small compared to intrinsic red noise in all pulsars, the distribution on $\hat A^2_{\textrm{gw}}$ in \autoref{eq:a_gw_estimator_no_correlations} is approximately Gaussian with a variance given by \autoref{eq:sigma_gw_estimator_no_correlations} except in the tails~\cite{Hazboun:2023tiq}.
In present analyses, $A^2_{\textrm{gw}}$ is not smaller than the corresponding intrinsic red noise amplitude for at least a few pulsars~\cite{NG15.GWB,NG15.detchar}. Therefore, to test how well the small signal approximation works, we perform 200 simulations with the length and cadence of the data set given in~\cite{NG11.gwb} and with red noise drawn from $p(\vect\eta | \vect{\delta t})$ from~\cite{NG12.gwb}.
We draw $A_{\textrm{gw}}$ from a uniform distribution, $\log_{10} A_{\textrm{gw, inj}}\in [-17, -13]$.
On each simulated dataset we calculate $\hat A^2_{\textrm{gw}}$ and its variance using \autoref{eq:a_gw_estimator_no_correlations} and \autoref{eq:sigma_gw_estimator_no_correlations}, and we use the method described in Appendix~\ref{app:pp_plot_interpretation} to construct PP-like plots, replacing the posterior samples with a Gaussian, $\mathcal N(\hat A_{\textrm{gw}}^2, \sigma^2_{\textrm{gw}})$. These are not traditional PP-plots, as there is no well-defined prior from which we draw our simulations and sample our posterior. However, performing the same processing as in Appendix~\ref{app:pp_plot_interpretation} does test how frequently the Gaussian distribution centered on $\hat A^2_{\textrm{gw}}$ with variance $\sigma_{\textrm{gw}}^2$ includes the simulated value of $A^2_\textrm{gw}$.  We do the same thing for the estimator that includes covariances between pulsar pairs, defined in \autoref{eq:estimator_with_correlations} and \autoref{eq:sigma_gw_with_correlations}.

\begin{figure}
    \centering
    \includegraphics[width=0.49\textwidth]{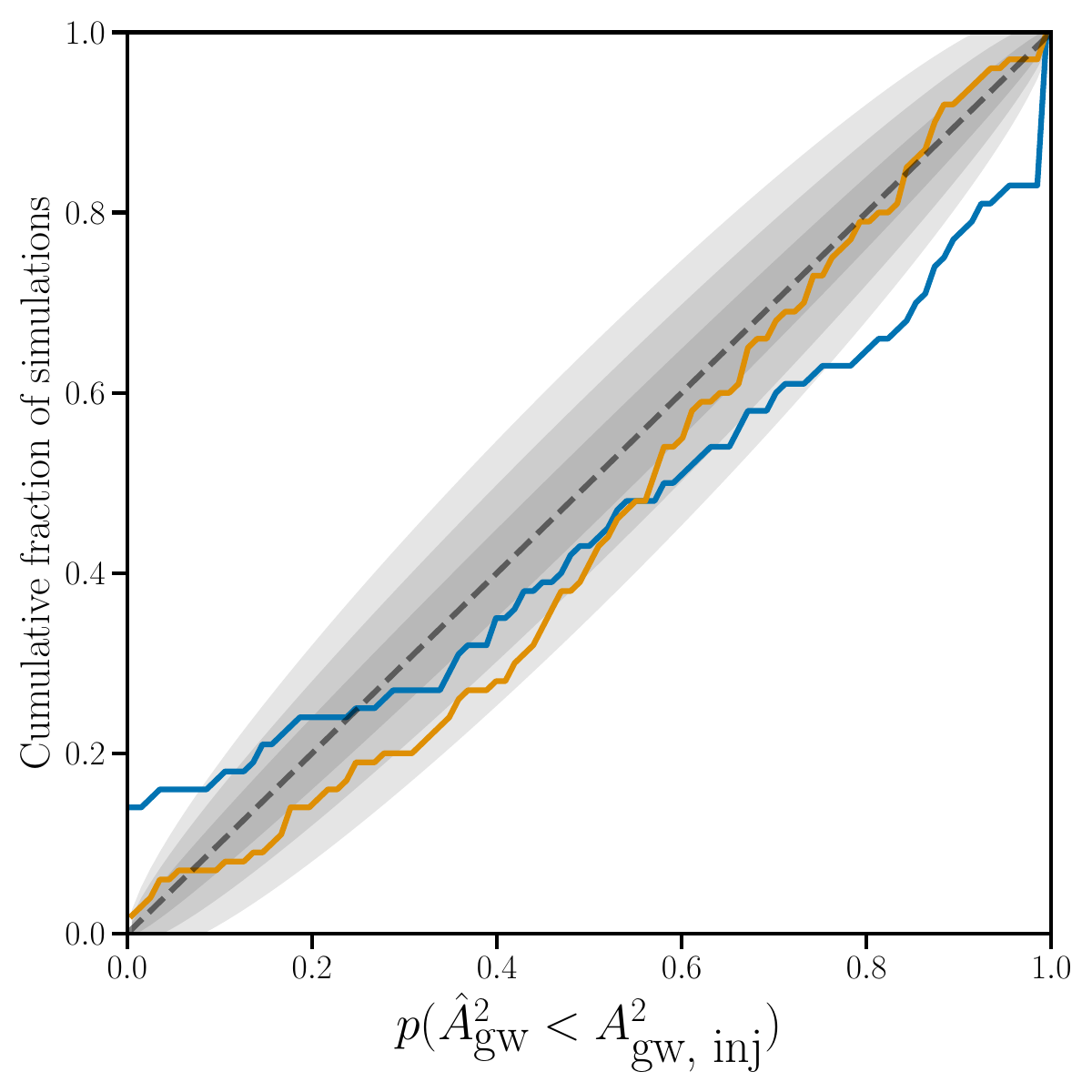}
    \caption{P-P-like plot that characterizes how well the optimal statistic functions as an estimator for the amplitude of the GWB. The blue, solid curve, which uses the traditional OS does not follow the expected diagonal line, indicating that it underestimates the variance on the estimator for the background. The orange curve, meanwhile, follows the diagonal line with its expected error bars (shaded region) because it properly estimates the variance by including the contribution from the GWB.}
    \label{fig:os_pp_plot_full}
\end{figure}

The results of this test are shown in \autoref{fig:os_pp_plot_full}, with the ``traditional'' optimal statistic results shown in the blue, solid curve and the corrected results shown in the orange, solid curve. The blue curve is consistent with the traditional optimal statistic underestimating the error on the estimator by not accounting for the GWB, and therefore not capturing the simulated value in its credible intervals as frequently as it should. The orange curve corrects this, as it follows the expected line more closely. Therefore, if one is to use the optimal statistic as an estimator for the GWB, the corrected statistic performs significantly better.

\begin{figure}
    \centering
    \includegraphics[width=0.49\textwidth]{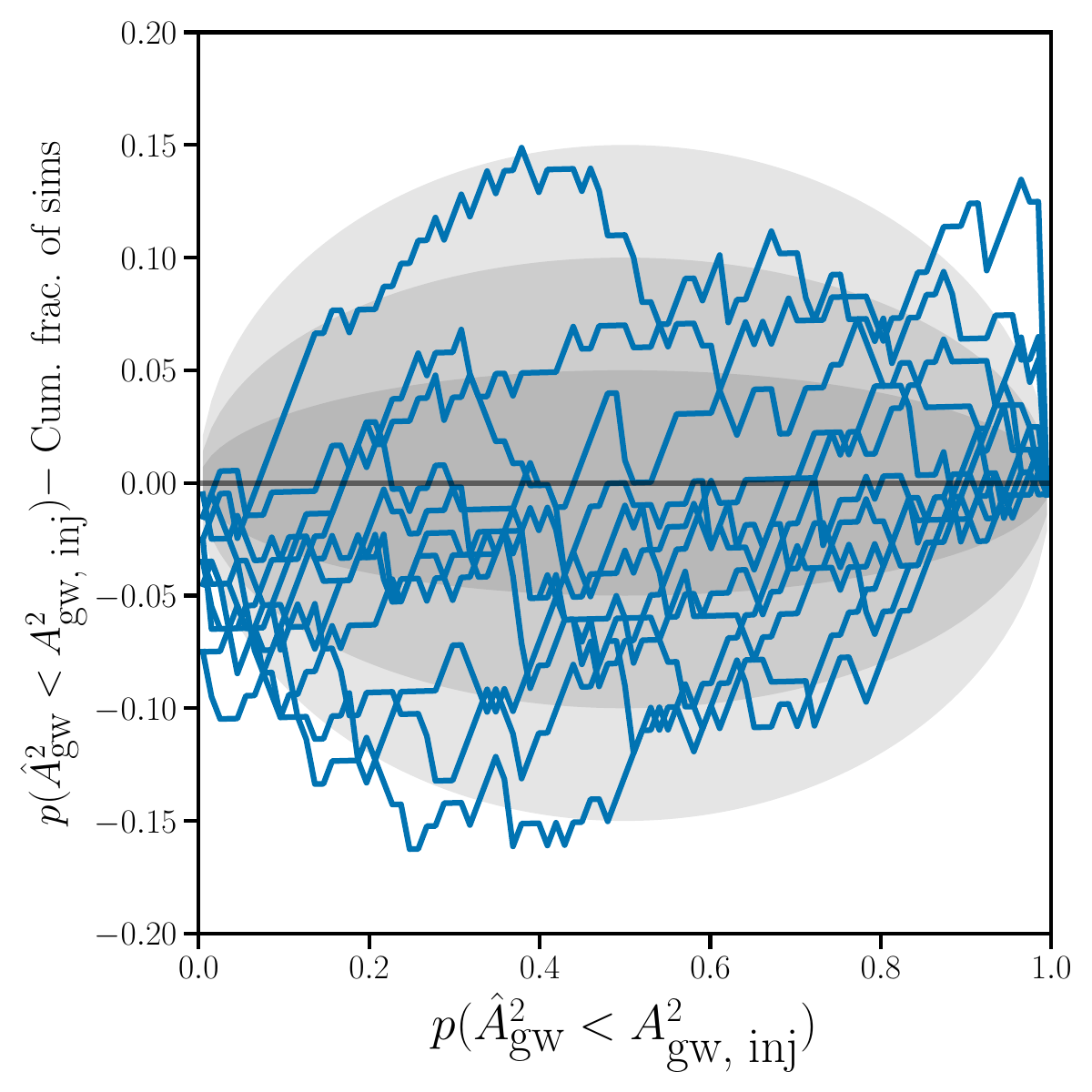}
    \includegraphics[width=0.49\textwidth]{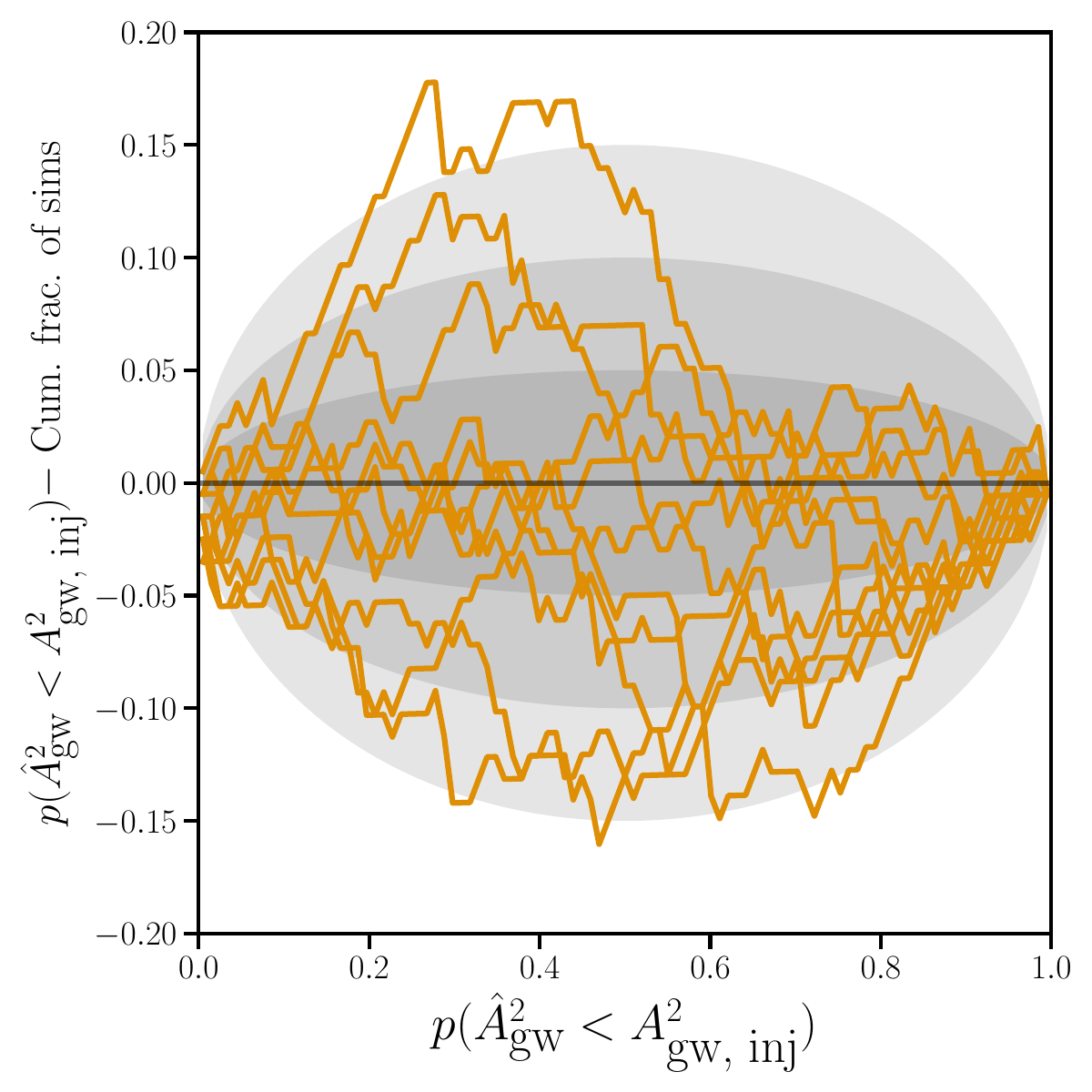}
    \caption{P-P-like plots (with diagonal subtracted off) that characterize how well the angular-binned optimal statistic functions as an estimator for the amplitude of the GWB, modulated by the Hellings-Downs correlations. We show results estimating the amplitude in each of 11 individual bins. Top: the blue, solid curve, uses the traditional binned OS does not follow the expected horizontal line, indicating that it underestimates the variance on the estimator for the background. This is especially obvious looking near zero and near one, where we see the curves diverge from the expected range (shaded regions) Bottom: the orange curve follows the horizontal line better than the blue curves in the top, because it properly estimates the variance by including the contribution from the GWB.}
    \label{fig:os_pp_plot_binned}
\end{figure}

We can perform the same test for the binned optimal statistic in \autoref{eq:binned_estimator_with_covariances} and \autoref{eq:binned_estimator_covariance_matrix} as well. For each simulation, we evaluate the cumulative distribution function at the simulated value, similar to the test in Appendix~\ref{app:pp_plot_interpretation}, but on each individual binned estimator in \autoref{eq:binned_estimator_with_covariances} and \autoref{eq:binned_estimator_covariance_matrix}. We plot the results of the P-P-like test in \autoref{fig:os_pp_plot_binned}, where we see similar results to \autoref{fig:os_pp_plot_full}. In this case, we show \textit{deviations} from the predicted line, and so expectation is zero, as opposed to $y=x$. The blue curves show excess cases where the CDF is zero and close to one, indicating misestimation. The orange curves, which includes the GWB in its variance, perform better, especially near zero and one. The orange curves do reach the 3-sigma level more than one might normally expect, and this is likely because we have assumed that the distribution on the binned estimators is a Gaussian. In practice, the distribution on the estimators is a generalized chi-squared distribution, which can be well approximated by a Gaussian under certain circumstances~\cite{Hazboun:2023tiq}.

In these simulations we have used the simulated amplitude of the background when calculating the covariance between correlations. We do this for practical purposes--the goal of these P-P-like tests is to show that our estimator is unbiased and its error bars are correct when we take into account the amplitude of the GWB. In practice, we do not have access to the GWB amplitude \textit{a priori}--we either use values drawn from a MCMC chain (e.g. the noise-marginalized optimal statistic), or we could employ an iterative approach, where we calculate the GWB estimator using the optimal statistic, and then use the estimator in the covariance matrix to properly estimate the covariance matrix from the correlations, and then repeat until convergence.

\subsection{The optimal statistic for detection}
We have shown that the estimators which include the strength of the GWB in their construction are better than the ones that do not, especially in the case where the amplitude of the GWB is of a similar size to that of the intrinsic red noise. However, the S/N calculated in~\autoref{eq:snr_os_no_correlations} is calculated under the null hypothesis. Therefore, when making a detection, we construct a distribution for this statistic under the null hypothesis (i.e., no correlated power), and then we compare the \textit{same statistic} calculated on the original data to that distribution. Therefore, \autoref{eq:snr_os_no_correlations} is the correct expression for a detection statistic. It is common to use a noise-marginalized version of this statistic, as discussed previously. That is, we take the average of this S/N over many draws from $p(\vect\eta | \vect{\delta t})$, and compare this to a null distribution.

Construction of a distribution for the null statistic takes a few forms. The analytic distribution of this statistic was calculated in~\cite{Hazboun:2023tiq}, and can be used. However, pathologies in the data that are not correctly modeled would not be accounted for in an analytic calculation. To preserve potential mismodeling, but still approximate the null distribution of the detection statistic, it is common to use sky scrambling~\cite{Cornish:2015ikx} and phase shifting~\cite{Taylor:2016gpq}. Sky scrambling involves assigning random sky positions to each pulsar, and calculating the optimal statistic, while phase shifting applies random phases to each frequency and each pulsar in the $\vect F$-matrix used to construct~\autoref{eq:s_matrix_os}. Comparisons between the analytic distribution and distributions constructed by sky scrambling and phase shifting are still in progress. Additionally, some concerns have been raised about whether performing sky scrambling and phase shifting without producing independent\footnote{A statistic one could use to assess independence is presented in~\cite{DiMarco:2023jqq}} scrambles or shifts results in oversampling parts of the null distribution, and undersampling other parts~\cite{DiMarco:2023jqq}. Some tests of this have been done to explore this in~\cite{NG15.GWB}, and found that more stringent ``orthogonality'' conditions do not produce meaningfully different distributions than those calculated with less stringent conditions. However, more tests are in progress.

\subsection{MCOS Tests}
We do not present novel tests of the MCOS here, as tests have been presented in separate work that more completely presents the statistic itself and characterizes its behavior~\cite{sardesai_generalized_2023}. We summarize those results here, for completeness.

In~\cite{sardesai_generalized_2023} and \cite{NG12.altpol} it is shown that, due to the non-isotropic distribution of pulsars on the sky, Hellings-Downs correlations are not orthogonal to monopolar or dipolar correlations, as one might expect. This is what motivates the construction of a statistic that simultaneously fits for multiple spatial correlation patterns simultaneously. In~\cite{sardesai_generalized_2023}, they perform simulations of non-Hellings-Downs correlation patterns, and show that it is possible using the statistic in~\autoref{eq:snr_os_no_correlations} to find a spurious detection of Hellings-Downs correlations. Using~\autoref{eq:mcos_estimator} and \autoref{eq:mcos_estimator_sigma}, remedies this situation, as the new statistic correctly finds that the non-Hellings-Downs correlations are preferred. Additionally, the authors consider the MCOS as an estimator for the strength of the background associated with each spatial correlation pattern. They find that, when multiple correlation patterns are simulated into the data, the MCOS estimator for the strength of each pattern performs better than individually estimating the strength of each process separately. 
However, there are issues with the estimation of signals and their uncertainties owing to the fact that the MCOS is based on the 
traditional optimal statistic, which does not take into account covariance between pulsar pairs (see \autoref{subsec:optstat_est}).

\section{Conclusions}
\label{sec:concl}

Here we have checked both Bayesian and frequentist methods used in the 15-year GWB analysis.
These methods were outlined as they are used in the 15-year analysis.
We subjected the method implementations of each part of the analysis to the most stringent tests of correctness performed so far, and we find that each analysis returns unbiased, self-consistent results.

In the Bayesian tests, we find that the priors are recovered correctly while including parallel tempering, AM, SCAM, DE, empirical distributions, and prior proposal distributions that are used in the full analysis.
This indicates that all proposal distributions work properly and are not biasing our results.
We also find that the simulations made with simulations pulled from a prior distribution return diagonal P-P plots within acceptable uncertainties.
By creating realizations of the 15-year data with various amplitudes and spectral indices of a GWB including HD correlations in them, we perform tests using reweighting, nested sampling, and the hypermodel to return Bayes factors between an HD correlated and CURN model.
Each comparison between these methods return log Bayes factor ratios consistent with zero.
Finally, we perform model comparison between a model and itself on these same simulations and find the expected result of Gaussian distributed Bayes factors centered around unity.

In the frequentist tests, we show through simulating data sets that properly including the GWB when constructing the optimal statistic is necessary in the situation where the GWB is not small compared to the intrinsic red noise. The estimator that accounts for the size of the GWB, however, is not consistent with the null hypothesis of no correlated power in the timing residuals, and therefore the ``traditional'' optimal statistic should be used as a detection statistic. Such a detection statistic should be calculated on the observed timing residuals and compared to a null distribution, which can be calculated either using the analytic distribution~\cite{Hazboun:2023tiq} or a method that preserves potential mismodeling but suppresses correlations~\cite{Cornish:2015ikx,Taylor:2016gpq}. Finally, we summarize the recently-proposed MCOS, which simultaneously fits for multiple spatial correlation patterns, and helps prevent, e.g., monopolar correlations from producing a spurious detection of Hellings-Downs correlations and vice-versa. 

Currently, the NANOGrav software collection allows one to perform inference without a steep learning curve.
As our data volume and parameter space increases with each data set, we look for additional methods to improve efficiency of memory management and the likelihood evaluation speed in \texttt{Enterprise}.
This includes possible data compression techniques to reduce the number of points of data required to fit to perform inference (e.g, \cite{gibbs_paper}).
Additionally, reducing the auto-correlation of our chains produced with \texttt{PTMCMCSampler} could reduce computation time significantly.
To this end, some recent work has used \texttt{JAX}~\cite{jax2018github}, a Python package that includes just-in-time compilation (JIT) to speed up evaluation of loops and auto-differentiation to take derivatives quickly and accurately, among other convenient features.
With this, \citet{freedman_efficient_2023} were able to use a No-U-Turn sampler (NUTs), a Hamiltonian Monte Carlo (HMC) sampling technique, which reduced auto-correlation significantly compared to \texttt{PTMCMCSampler}.
In \citet{becsy_fast_2022}, JIT is used through the \texttt{numba} Python package alongside techniques to sample some parameters more often than others to vastly increase the speed of evaluation of continuous GW searches.
While these papers have shown the incredible speed gains that one may achieve via these new technologies, they cannot be easily implemented in the current paradigm used by \texttt{Enterprise}.
This is primarily due to sub-classing of \texttt{numpy} arrays used in \texttt{Enterprise}, which is not supported by these other programs.

Both the Bayesian and frequentist methods and their software representations pass all the tests they were subjected to.
As PTA data sets increase in sensitivity, testing that our methods are reliable proves paramount.
Here, we validate the results of current and future analyses that use the methods examined.
Through rigorous testing of our software and methods, we form a foundation on which astrophysical results may stand.

\begin{acknowledgments}
The NANOGrav project receives support from National Science Foundation (NSF) Physics Frontiers Center award number 1430284.
A.D.J., K.C., and M.V. acknowledge support from the Caltech and Jet Propulsion Laboratory President's and Director's Research and Development Fund.
A.D.J. and K.C. acknowledge support from the Sloan Foundation.
S.H. is supported by the National Science Foundation Graduate Research Fellowship under Grant No. DGE-1745301.
L.B. acknowledges support from the National Science Foundation under award AST-1909933 and from the Research Corporation for Science Advancement under Cottrell Scholar Award No. 27553.
P.R.B. is supported by the Science and Technology Facilities Council, grant number ST/W000946/1.
S.B. gratefully acknowledges the support of a Sloan Fellowship, and the support of NSF under award \#1815664.
M.C. and S.R.T. acknowledge support from NSF AST-2007993.
M.C. and N.S.P. were supported by the Vanderbilt Initiative in Data Intensive Astrophysics (VIDA) Fellowship.
Support for this work was provided by the NSF through the Grote Reber Fellowship Program administered by Associated Universities, Inc./National Radio Astronomy Observatory.
Support for H.T.C. is provided by NASA through the NASA Hubble Fellowship Program grant \#HST-HF2-51453.001 awarded by the Space Telescope Science Institute, which is operated by the Association of Universities for Research in Astronomy, Inc., for NASA, under contract NAS5-26555.
M.E.D. acknowledges support from the Naval Research Laboratory by NASA under contract S-15633Y.
T.D. and M.T.L. are supported by an NSF Astronomy and Astrophysics Grant (AAG) award number 2009468.
E.C.F. is supported by NASA under award number 80GSFC21M0002.
G.E.F., S.C.S., and S.J.V. are supported by NSF award PHY-2011772.
The Flatiron Institute is supported by the Simons Foundation.
The work of N.La. and X.S. is partly supported by the George and Hannah Bolinger Memorial Fund in the College of Science at Oregon State University.
N.La. acknowledges the support from Larry W. Martin and Joyce B. O'Neill Endowed Fellowship in the College of Science at Oregon State University.
Part of this research was carried out at the Jet Propulsion Laboratory, California Institute of Technology, under a contract with the National Aeronautics and Space Administration (80NM0018D0004).
M.A.M. is supported by NSF \#1458952 and \#2009425.
C.M.F.M. was supported in part by the National Science Foundation under Grants No. NSF PHY-1748958 and AST-2106552.
A.Mi. is supported by the Deutsche Forschungsgemeinschaft under Germany's Excellence Strategy - EXC 2121 Quantum Universe - 390833306.
K.D.O. was supported in part by NSF Grant No. 2207267.
T.T.P. acknowledges support from the Extragalactic Astrophysics Research Group at E\"{o}tv\"{o}s Lor\'{a}nd University, funded by the E\"{o}tv\"{o}s Lor\'{a}nd Research Network (ELKH), which was used during the development of this research.
S.M.R. and I.H.S. are CIFAR Fellows.
Portions of this work performed at NRL were supported by ONR 6.1 basic research funding.
J.D.R. also acknowledges support from start-up funds from Texas Tech University.
J.S. is supported by an NSF Astronomy and Astrophysics Postdoctoral Fellowship under award AST-2202388, and acknowledges previous support by the NSF under award 1847938.
Pulsar research at UBC is supported by an NSERC Discovery Grant and by CIFAR.
S.R.T. acknowledges support from an NSF CAREER award \#2146016.
C.U. acknowledges support from BGU (Kreitman fellowship), and the Council for Higher Education and Israel Academy of Sciences and Humanities (Excellence fellowship).
C.A.W. acknowledges support from CIERA, the Adler Planetarium, and the Brinson Foundation through a CIERA-Adler postdoctoral fellowship.
O.Y. is supported by the National Science Foundation Graduate Research Fellowship under Grant No. DGE-2139292.
J.A.C.C. was supported in part by NASA CT Space Grant PTE Federal Award Number 80NSSC20M0129, and also supportedin part by the National Science Foundation's NANOGrav Physics Frontier Center, Award Number 2020265.
\end{acknowledgments}

\appendix
\section{Structure of \texttt{Enterprise}}
\label{app:struct_enterprise}
The Enhanced Numerical Toolbox Enabling a Robust PulsaR Inference SuitE (\texttt{Enterprise})~\cite{enterprise} is a Python package built to analyze pulsar noise and timing models, and to search for GWs in PTA data.
It grew out of previous PTA data analysis packages such as \texttt{NX01}~\cite{taylor_nx01_2016}, \texttt{PAL2}~\cite{justin_ellis_2017_251456}, and \texttt{piccard}~\cite{van_haasteren_piccard_2016}.
In the context of this paper, we are using \texttt{Enterprise} to search for an isotropic GWB in simulations of the NANOGrav 15-year data set.
Development of \texttt{Enterprise} began in the interest of making a suite of tools that could analyze data from any PTA consortium, and support international collaboration without requiring significant knowledge of programming.

\begin{figure*}
\includegraphics[width=1.5\columnwidth]{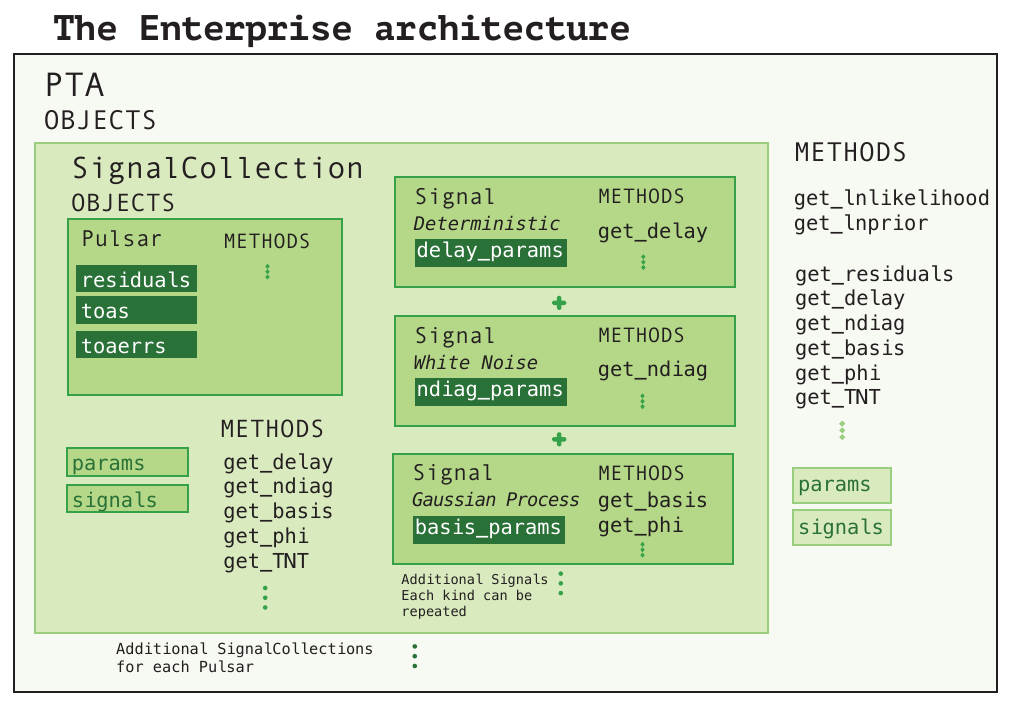}
\caption{The hierarchical structure of \texttt{Enterprise}. \label{fig:enterprise}}
\end{figure*}

\texttt{Enterprise} is a thoroughly object-oriented package that defines a PTA data model as a hierarchical structure (see  \autoref{fig:enterprise}).
The top-level object is \texttt{PTA}, which interfaces with the sampler by way of the \texttt{get\_lnlikelihood} and \texttt{get\_lnprior} methods: these evaluate the (log) Gaussian-process-marginalized likelihood of \autoref{eq:trad_like} and the total prior respectively, taking as input a Python dictionary of parameter values, or alternatively a \texttt{numpy} vector with the parameters in the same order as in the property \texttt{PTA.params}.

The \texttt{PTA} object is created from a sequence of \texttt{SignalCollection} objects, each of which corresponds to a pulsar in the array, and represents the pulsar's complete data model. \texttt{SignalCollection} implements a set of methods that return the vector and matrix constituents used by \texttt{PTA.get\_lnlikelihood}: for instance, \texttt{get\_residuals} for the residuals $\vect{\delta t}$, \texttt{get\_delay} for any deterministic delays, \texttt{get\_ndiag} for the white-noise matrix $\vect{N}$, \texttt{get\_basis} and \texttt{get\_phi} for Gaussian-process bases $\vect{T}$ and prior matrix $\vect{B}$. All these methods take a Python dictionary of parameter values.

Each \texttt{SignalCollection} object consists of one \texttt{Pulsar} object and any number of \texttt{Signal} objects. \texttt{Pulsar} uses the \texttt{PINT} or \texttt{libstempo} packages to read pulsar data from \texttt{par} and \texttt{tim} files, standard formats for storing timing model parameters and timing data (see, e.g., \cite{luo_pint_2021}), and it provides the vectors \texttt{residuals} (for $\vect{\delta t}$), \texttt{toaerrs} (for $\vect{\sigma}$), and \texttt{freqs} (for the pulse radio frequencies), as well as the timing-model design matrix \texttt{Mmat} (i.e., $\vect{M}$), and several more pulsar properties. The \texttt{Signal} objects, which come in a variety of subtypes, implement deterministic signals and noise components, and provide the building blocks that \texttt{SignalCollection} and (downstream) \texttt{PTA} assemble into a full likelihood. Specifically, deterministic signals define \texttt{get\_delay}, white-noise components define \texttt{get\_ndiag}, and Gaussian processes define \texttt{get\_basis} and \texttt{get\_phi}.
\emph{Common} Gaussian processes, used most notably for the HD-correlated background, have parallel bases for each pulsar (with the same $n_\mathrm{GP}$ but different $n_\mathrm{obs}$), provide a \texttt{get\_phicross} method in addition to \texttt{get\_phi} to fill off-diagonal prior-matrix elements, and are meant to be used with common hyperparameters.

The package includes a number of optimizations for speed, memory, and ease of configuration:
\begin{itemize}
    \item All objects keep track of which parameters affect the output of their methods, and cache results if those parameters are not changed, or are set as constants. This is useful, for instance, in analyses that freeze the white-noise parameters, or in stochastic schemes that update parameters in blocks. The \texttt{SignalCollection} and \texttt{PTA} objects have methods for intermediate matrix combinations such as $\vect{T}^T \vect{N}^{-1} \vect{T}$, and these are also cached on the basis of the relevant parameters. 
    \item The \texttt{SignalCollection} object combines delays from all deterministic signals into a single vector, assembles white-noise matrices into a single matrix, and stacks Gaussian-process bases and priors into two combined matrices. The object has the ability of reusing basis vectors that appear identically in multiple processes, such as Fourier vectors for red noise and the GWB.
    \item The \texttt{get\_ndiag} methods return custom ``kernel'' objects that represent the constituents of the measurement-noise matrix $\vect{N}$. Whether the kernels are represented internally as a vector for EFAC/EQUAD noise, a Sherman-Morrison decomposition for ECORR [see \autoref{eq:shmo}], or a sparse matrix for even more general cases, they all provide ``solve'' methods that return combinations such as $\vect{T}^T \vect{N}^{-1} \vect{y}$ and $\vect{T}^T \vect{N}^{-1} \vect{T}$ without actually computing and storing $\vect{N}^{-1}$. Furthermore, the kernel objects know how to combine themselves with other kernel objects, creating an optimized object used in the likelihood calculation.
    \item The \texttt{PTA} object defines an optimized \texttt{get\_phiinv} method that can choose between multiple inversion strategies to build the global $\vect{B}^{-1}$ depending on the structure of $\vect{B}$.
\end{itemize}

\texttt{Enterprise} is configured by building a \texttt{SignalCollection} \emph{template} from a sequence of \texttt{Signal} templates. \texttt{Enterprise} includes ``factories'' that build templates for commonly used noise components such as \texttt{MeasurementNoise} (for radiometer noise), \texttt{EcorrKernelNoise} (for jitter-like noise), \texttt{TimingModel} (for the Gaussian process with improper prior used to marginalize over timing-model corrections), \texttt{FourierBasisGP} (for Gaussian processes with a sine/cosine Fourier basis), and \texttt{FourierBasisCommonGP} (for Fourier Gaussian processes with correlations across pulsars), and more.
Each template must be assigned one or more \texttt{Parameter} objects that encode the priors chosen for the relevant parameters. For instance, \texttt{MeasurementNoise} may be passed 
\texttt{efac = Normal(1, 0.25)} and
\texttt{log10\_t2equad = Uniform(-8.5, -5)} to indicate that in \autoref{eq:whitenoise} $F \sim \mathcal{N}(1, 0.25)$ and $\log_{10} Q \sim U(-8, -5)$. Parameters may also be passed as \texttt{Constant} if they will not be sampled stochastically.

The \texttt{SignalCollection} template is then applied to each \texttt{Pulsar} object in turn, generating \emph{instantiated} \texttt{SignalCollection} and \texttt{Signal} objects, in which parameters are specialized to the pulsar (e.g., \texttt{efac} becomes \texttt{B1855+09\_efac}) and the \texttt{Pulsar} quantities are made available locally.
For most \texttt{Signal}s, a further kind of specialization is possible in which a \emph{selection} function is passed to the \texttt{Signal}, effectively splitting it into multiple copies, each applied with different parameters to a different section of the data. For instance, the selection \texttt{by\_backend}, when given to \texttt{MeasurementNoise}, would split EFAC parameters for the NANOGrav B1855+09 pulsar into 
\texttt{B1855+09\_430\_ASP\_efac},
\texttt{B1855+09\_430\_PUPPI\_efac},
\texttt{B1855+09\_L-wide\_ASP\_efac},
\texttt{B1855+09\_L-wide\_PUPPI\_efac},
each applied to the subset of the residuals obtained with that receiver backend.

Configuration is completed by collecting the set of instantiated \texttt{SignalCollection} into a single \texttt{PTA} object. See the NANOGrav 12.5-yr and 15-yr tutorials for examples of creating a standard \texttt{PTA} object that is ready to compute log likelihoods and priors.

\texttt{Enterprise} includes a number of additional facilities and utilities that simplify the task of creating a model.
For instance, the prior for a Fourier Gaussian processes can be written simply as Python functions that take a vector of frequencies and return the diagonal components of $\vect{\phi}$. By wrapping the prior in \texttt{parameter.function}, the arguments of the Python function are automatically book-kept as model parameters and specialized to pulsar and selections. Furthermore, if the Python function includes arguments that are defined in the \texttt{Pulsar} object (such as \texttt{residuals} or \texttt{toaerrs}), these are passed to the function automatically once the \texttt{Signal} that uses the function has been specified.

Of course, \texttt{Enterprise} already defines a number of commonly used Gaussian-process priors and ORFs. It also implements the deterministic model of solar-system-ephemeris uncertainties \cite{vallisneri_modeling_2020} used for older NANOGrav data releases  (\texttt{utils.FourierBasisCommonGP\_physicalephem}); it can draw random values of model parameters according to their prior (\texttt{parameter.sample}); it can sample the conditional distributions of Gaussian-process weights given their hyperparameters (\texttt{utils.ConditionalGP}); it can simulate full realizations of the data model (\texttt{simulate}); it can handle the faster likelihood of \autoref{sec:fasterlike} with a special noise kernel \texttt{MarginalizingTimingModel} that holds the components of $\vect{D}$ internally and performs the Woodbury inversion of \autoref{eq:fastwoodbury} transparently.
See the \texttt{Enterprise} documentation for this and much more.

\section{Asynchronous Parallel Tempering and the HyperModel Framework}
\label{sec:syncsampler}
\texttt{PTMCMCSampler} previously used an asynchronous parallelization model for parallel tempering in which each chain sampled a set number of steps on its own unless interrupted by another chain asking to propose a swap between chains.
During our tests, we noticed that this asynchronous 
parallel-tempered MCMC biased Bayes factors computed with hypermodel in favor of the model that took the longest evaluation time per iteration.
As a simple example of this issue,
we simulate data with a sinusoidal signal $h(t)$ and noise $n(t)$,
\begin{equation}
    \textbf{d}(t) = \textbf{h}(t) + \textbf{n}(t)\,.
\end{equation}
The sinusoid signal is described by amplitude, angular frequency, and phase,
\begin{equation}
    h_i(t, A, \omega, \phi) = A \sin(\omega t_i + \phi)\,,
\end{equation}
and the noise $n(t) \sim \mathcal{N}(0, 1)$.
The log likelihood we use for this signal model is
\begin{equation}
    p(d \mid A, \omega, \phi) \propto -\frac{1}{2} \sum_i \left(d_i(t) - h_i(t, A, \omega, \phi)\right)^2\,.
\end{equation}
Priors are all chosen to be uniform with 
\begin{align}
    A&\sim\text{Uniform}[0, 5]\,,\\
    \omega&\sim \text{Uniform}[0, 3]\,,\\
    \phi&\sim \text{Uniform}[0, \pi]\,.
\end{align}
Priors remain the same across each model used, and parameter estimation of the simulated values returns consistently with the posteriors.

The \texttt{HyperModel} framework consists of multiple models concatenated with a continuous ``switch'' parameter between them.
This framework chooses models based on which of two bins the switch parameter falls into.
In this case, the first model is the sinusoidal model as described above, and the second model is the same model, but with a \texttt{time.sleep()} call to increase the evaluation time by a factor of a few.

\begin{figure}
    \centering
   \includegraphics[width=\columnwidth]{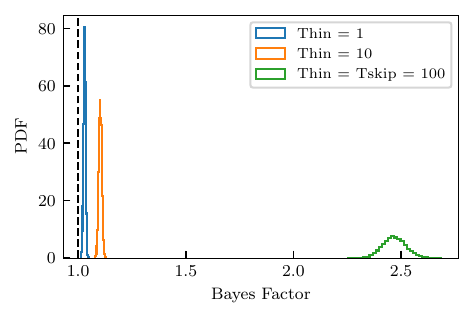}
    \caption{The Bayes factor for a model over itself should be one, which is shown as a dashed, black line on this plot.
    Instead, we find that the Bayes factor is inconsistent with a value of one.
    Parallel tempering swaps are proposed every 100 samples, labeled ``Tskip'' in this plot (as it is in \texttt{PTMCMCSampler}).
    Thinning increases the contamination from swaps by increasing the number of samples that come from swaps out of the total number.
    Uncertainties on the Bayes factors were computed via bootstrapping.}
    \label{fig:bf_prob}
\end{figure}

Upon computing the odds ratios with the asynchronously updated sampler, we find that the odds ratio is not consistent with the anticipated value of one, as seen in \autoref{fig:bf_prob}.
Parallel tempering swap proposals occur every 100 samples.
Thinning by multiples of 10 cause the samples to become increasingly contaminated with swaps that pull the odds ratio away from one.

\begin{figure*}
    \centering
    \includegraphics[width=\textwidth]{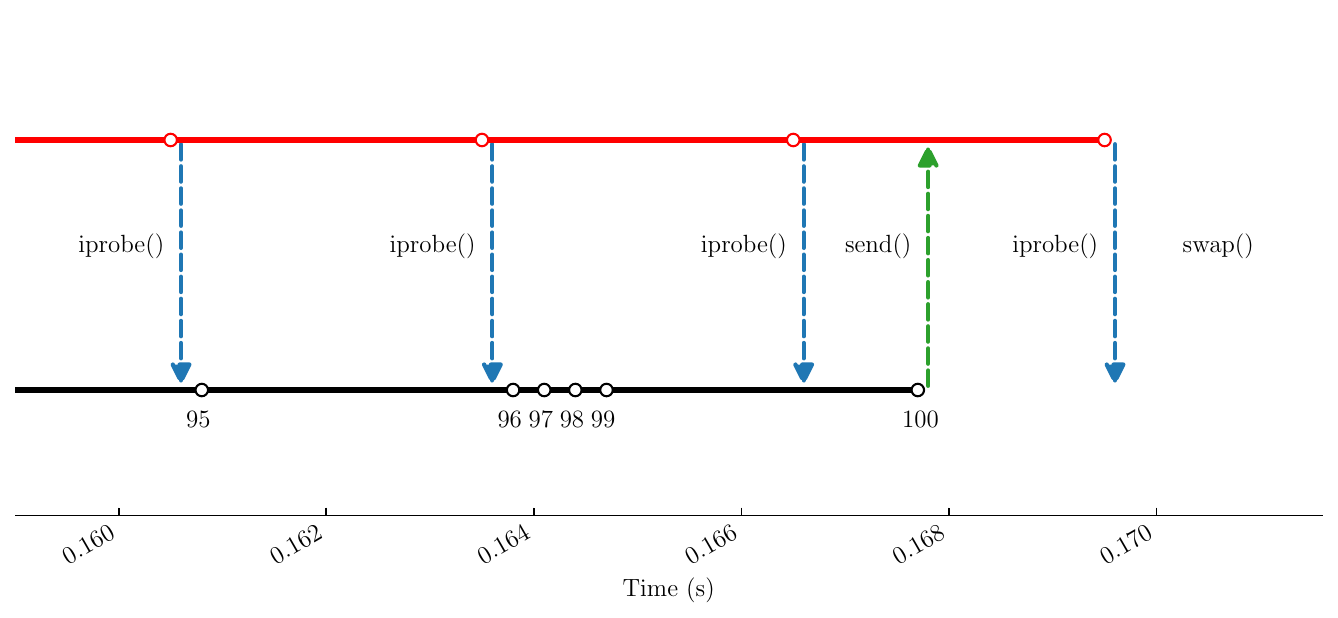}
    \caption{Two chains, one at $T_1 = 1$ and the other at $T_1 = 2$, progress through iterations. Iterations start at the white circles and last until the next circle. Two models are being considered, one of which takes 10 times longer to evaluate. After each iteration, the hot chain, shown in red on top, runs a non-blocking probe to check if anything has been sent from the cold chain, shown in black on bottom. Once the cold chain reaches 100 samples, a blocking send pushes data required for a swap to the hot chain. This data gets accepted by the hot chain after the current iteration finishes and a swap is proposed. Because of this disparity in evaluation times, we find that the hot chain is 10 times more likely to be in the model that takes longer to evaluate. This results in a bias in the model that goes into the swap proposals. Fundamentally, this means that the swaps proposed are more likely to happen in certain parts of the \texttt{HyperModel} parameter space, breaking detailed balance.}
    \label{fig:timeline}
\end{figure*}

Syncing the model evaluation times gives odds ratios consistent with one, contrary to what we see in \autoref{fig:bf_prob}
Therefore, the problem appears to be caused by the evaluation time of one of the models being much longer than the other.
The exact timeline of when these swaps are proposed is shown in \autoref{fig:timeline} and proves critical to figuring out what went wrong.

In asynchronous parallel tempering, swaps are proposed whenever a chain gets to a set number of samples.
The chain sends data to the next chain up in the temperature ladder, halting the chain until the other chain is done with its current sample evaluation.
While waiting for this signal, the hotter chain continues collecting samples, probing the lower chain for data after each sample.
Once the signal has been received, the chains swap with their most recent sample iteration.

Upon proposing a swap, the cold chain finds the hot chain in the model that has the longer evaluation time more often than not.
This means that the swaps are proposed with a dependence on where in the switch parameter space we are, violating detailed balance for parallel tempering swap proposals.

One option to fix this involves synchronizing the swap proposals so that all chains propose swaps at the same iterations.
The other option is to weight the proposals with weights proportional to the ratio of the evaluation times.
In some situations this option is not possible due to dependence of evaluation time on where we are in the parameter space.
We opted to synchronize the sampler for simplicity and to keep the sampler as generic as possible.

\section{Understanding Confidence Intervals of P-P Plots}
\label{app:pp_plot_interpretation}
In \autoref{fig:pp_plot}, the confidence intervals are found as in~\citet{ashton_bilby-mcmc_2021}.
Unlike in the Q-Q plots where we have tens of thousands of samples, the standard bootstrap approximated confidence intervals (which estimates the confidence intervals as Gaussian) overshoot the exact bounds near p-values of 0 and 1.
For a given realization of the data, we have a simulated value for a single parameter $\theta_\text{inj}$, and we can compute a CDF of this value,
\begin{equation}
    F(\theta_\text{inj}) = \int_{-\infty}^{\theta_\text{inj}} p(x \mid \vect{\delta t}) dx.
\end{equation}
If we take many new realizations, their associated CDFs should be uniformly distributed between 0 and 1.
Picking a particular p-value on the horizontal axis will split the uniform distribution into two segments.
Let us define a success as $F(\theta_{inj}) \leq s$ and a failure as $F(\theta_{inj}) > s$ where $s$ is the chosen horizontal axis value.
From $N$ draws, the probability of $k$ successes can be found with a binomial distribution,
\begin{equation}
    p(k \text{ successes}) = \binom{N}{k} s^k(1-s)^{N - k}.
\end{equation}
We would like to have coverage $(1 - \alpha) / 2$ on both sides of the mean of the distribution with
\begin{equation}
    \alpha = (0.6827, 0.9545, 0.9973)
\end{equation}
for the three bounds that we show.
Using the quantile function for the binomial distribution, $F^{-1}(s)$ (\texttt{binom.ppf} in \texttt{scipy.stats}) on each of the horizontal axis values, the offset from the mean is
\begin{equation}
    \sigma = F^{-1}(s) / N.
\end{equation}

\section{Batch Updates of the Sample Mean and Covariance}
\label{app:online_alg}
Let $\vect{x_0}$ be an $N \times n_\text{param}$ matrix of samples from the history of an MCMC chain and let $\vect{x_1}$ be the next $M$ samples of the $n_\text{param}$ parameters.
Finally, let $\vect{x}$ be the concatenation of $\vect{x_0}$ and $\vect{x_1}$ with total length $N + M$.
The mean for $\vect{x}$ along the $N + M$ axis can be computed as
\begin{equation}
    \bar{x}^j = \bar{x}_0^j + \frac{1}{M + N} \sum_{i=1}^N \left(x_1^{ij} - \bar{x}_0^j\right),
\end{equation}
where an over bar denotes an average.
Next, we can compute the sample covariance through a batch update as
\begin{equation}
    \Sigma = \frac{(N - 1)\Sigma_0 + \Sigma_1}{N + M - 1},
\end{equation}
where
\begin{equation}
    \Sigma_1^{i j} = \left(x_1^{ji} - \bar{x}^j\right)\left(x_1^{ij} - \bar{x}_0^j\right).
\end{equation}
In \texttt{PTMCMCSampler}, these methods use $M = 1$ and iterate over the whole chain.

\bibliography{biblio}

\end{document}